\documentclass[preprint,12pt]{aastex61}

\received{2018 August 21}
\accepted{2018 December 9}
\submitjournal{ApJ}

\shorttitle{SHOCK-GENERATING PLANETESIMALS}
\shortauthors{NAGASAWA ET AL.}

\begin{document}
\title{ Shock-Generating Planetesimals Perturbed by \\ a Giant Planet in a Gas Disk}

\correspondingauthor{Makiko Nagasawa}
\email{nagasawa\_makiko@med.kurume-u.ac.jp}

\author{M. Nagasawa}
\affil{Department of Physics, Kurume University School of Medicine, 67 Asahi-machi, Kurume-city,
Fukuoka 830-0011, Japan}

\author{K. K. Tanaka}
\affil{Astronomical Institute, Tohoku University, 6-3 Aramaki, Aoba-ku Sendai, Miyagi 980-8578, Japan}

\author{H. Tanaka}
\affil{Astronomical Institute, Tohoku University, 6-3 Aramaki, Aoba-ku Sendai, Miyagi 980-8578, Japan}

\author{H. Nomura}
\affil{Department of Earth and Planetary Sciences, Tokyo Institute of Technology, 2-12-1, Ookayama, Meguro-ku, Tokyo 152-8551, Japan}

\author{T. Nakamoto}
\affil{Department of Earth and Planetary Sciences, Tokyo Institute of Technology, 2-12-1, Ookayama, Meguro-ku, Tokyo 152-8551, Japan}

\author{H. Miura}
\affil{Graduate School of Natural Sciences, Nagoya City University, Yamanohata 1, Mizuho-cho, Mizuho-ku, Nagoya 467-8501}

\begin{abstract}
We examined the excitations of planetesimals caused by the resonances of a giant planet in a protoplanetary gas disk.
The highly excited planetesimals generate bow shocks, the mechanism of which results in chondrule formation, crystallization of silicate dust, and evaporation of icy planetesimals.
The planetesimals beyond 2:1 resonance migrate owing to the gas drag and obtain the maximum eccentricity around 3:1 resonance, which is located at approximately half the planetary distance. 
The eccentricity depends on the parameters of the planetesimals and the Jovian planet, such as size and location, and gas density of the disk.
The maximum relative velocity of a 100-km-sized planetesimal with respect to the gas disk reaches up to $\sim 12$ ${\rm kms^{-1}}$ in the case of Jupiter owing to secular resonance, which occurs because of the disk's gravity. 
We find that if a Jovian mass planet is located within 10 au, the planetesimals larger than 100 km gain sufficient velocity to cause the melting of chondrule precursors and crystallization of the silicate. 
The maximum velocity is higher for large planetesimals and eccentric planets.
Planetesimals are trapped temporarily in the resonances and continue to have high speed over $\ga$1 Myr after the formation of a Jovian planet.
This duration fits into the timescale of chondrule formation suggested by the isotopic data.
The evaporation of icy planetesimals occurs when a Jovian planet is located within 15 au. 
This mechanism can be a new indicator of planet formation in exosystems if some molecules ejected from icy planetesimals are detected.
\end{abstract}
\keywords{meteorites, meteors, meteoroids --- minor planets, asteroids: general --- planets and satellites: dynamical evolution and stability --- planets and satellites: formation --- shock waves}

\section{INTRODUCTION}
\label{sec:intro}

\begin{figure}\epsscale{0.6}
\plotone{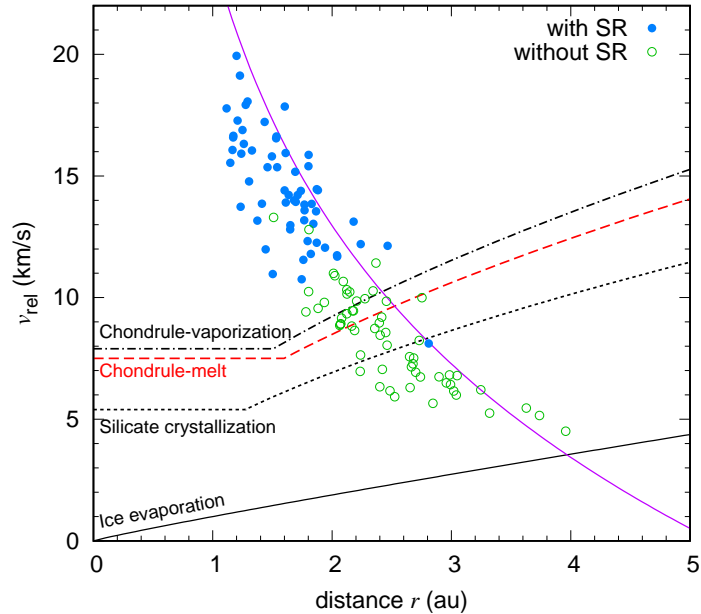}
\caption{ The required relative velocity of the planetesimals for chondrule melting (1900 K; red broken line), chondrule vaporization (2100 K; black dashed-dotted line), crystallization of amorphous silicate (1000 K; dotted line), and icy planetesimal evaporation (solid line) (see \S 4 for the details of the related equations).
Purple solid line shows the maximum relative velocity of the planetesimal whose apocenter crosses the Jovian orbit.
Minimum mass gas disk is assumed here for evaluating the shock temperature. 
Circles indicate the maximum relative velocities of sixty 300-km-sized planetesimals inside the Jovian orbit (similar simulations as performed by  Nagasawa et al. 2014).
Jovian secular resonance excites the planetesimal velocity when self-gravity of the gas disk is included (blue) as compared with the cases without the disk's gravity (green). 
Note that the circle indicates only the maximum relative velocity and its location. During the evolution, the planetesimal has relative velocity from 0 to the maximum value and travels across a wide region. 
(A colored version of this figure is available in the online journal.)
\label{fig:introvrel}}
\end{figure}

The observation of an asteroid belt shows depletion or piling-up of materials depending on the location of planet resonances. 
The resonances with Jupiter are especially noticeable. 
Asteroids between the Jovian 2:1 resonance (3.3 au) and Jupiter (5.2 au) have currently dispersed except for objects trapped in resonances (e.g., Trojans and Hildas).
Migration of asteroids (planetesimals) caused by gas drag accounts for the loss from the region beyond 2:1 resonance if the protoplanetary disk is preserved for $10^6-10^7$ yr after the formation of Jupiter (Ida \& Lin 1996; Marzari et al. 1997; Weidenschilling, Marzari, \& Hood 1998; Marzari\& Weidenschilling 2002).
During the migration, the planetesimals pass through many Jovian resonances, which excite the eccentricities of the planetesimals. 
These eccentricities are further boosted at the location of 2:1 resonance, and they normally reach as high as $e\sim 0.3-0.6$ at approximately 2--3 au, that is, they obtain a velocity of $v_{\rm rel} \la 8$ ${\rm km s^{-1}}$ relative to the gas disk (Marzari \& Weidenschilling 2002). 
Secular resonance is known to have a strong effect on the asteroidal orbit in the primordial gas disk (e.g., Ward 1981; Heppenheimer 1980; Lecar \& Franklin 1997). 
{ The secular resonances are related to the motions of the pericenter and the ascending nodes of the objects.
Periastra and nodes of objects rotate due to the gravity of other planets and the disk.
When the rotational speed of the periastron of an asteroid is close to the eigenfrequencies of the system, which is normally similar to the rotational speed of the periastra of planets, the eccentricity of the asteroid oscillates with large amplitude. 
It corresponds to the transfer of the angular momentum of the planet to the asteroid.
Similarly, when the rotational speed of the ascending node of an asteroid is close to the eigenfrequencies of the system, which is normally similar to the rotational speed of the ascending nodes of planets, it allows the inclination of the asteroid to oscillate with large amplitude. 
The inner boundary of the main belt of current asteroids ($\sim$ 2 au) is mainly determined by the secular resonances related to the rotation of the periastron and ascending node of Saturn caused by Jupiter. 
When the primordial gas disk exists, the eigenfrequencies of the system almost coincide with the rotational speeds of the periastra of planets caused by gravity of the disk.
In such a situation, the resonance related to Jupiter is most effective in the excitation of the eccentricities of the asteroids.}
The disk self-gravity should not be neglected when we consider the evolution of planetesimals experiencing gas drag. 
Nagasawa et al. (2014) showed that the Jovian secular resonance, which arises owing to the gas disk gravity, excites the 300-km-sized planetesimals up to $v_{\rm rel} \la 20$ ${\rm km s^{-1}}$ (Fig. \ref{fig:introvrel}).

The supersonic planetesimals, which appear following the formation of a Jovian planet, cause several interesting effects.
The shock wave generated by the planetesimals is well known as one of the formation mechanisms of chondrules, which are 0.1--1 mm-sized spherical grains commonly found in primitive meteorites. 
In the asteroid belt region of the minimum-mass disk ($\sim 10^{-10} {\rm g cm^{-3}}$), velocity of $\ga$ 10 ${\rm km s^{-1}}$ would be necessary for the complete melting of 1-mm-sized dust (Iida et al. 2001). 
{ In Figure \ref{fig:introvrel}, we show the conditions required to cause the melting of chondrule precursors (1900 K) and the conditions required for the initiation of vaporization of chondrules (2100 K) (see \S 4). 
In the outer region, the required velocities become higher because the gas density is small.
The maximum relative velocities of 300-km-sized planetesimals excited by Jupiter are indicated by circles.
To form chondrules, the maximum velocity should be higher than the melting-line.
Because the time interval of the peak-speed is not long and the planetesimal usually has a smaller velocity, it can contribute to the formation of chondrules even if the maximum speed exceeds the vaporization line.}
Jovian mean-motion resonances are proposed as the origin of the planetesimal bow shocks (e.g., Weidenschilling et al. 1998); however, secular resonance is essential to obtain such a high velocity (Nagasawa et al. 2014).
The planetesimal shock would also contribute to the formation of crystalline silicates in the disk; this silicate dust has been observed in protoplanetary disks (Henning 2010 and references therein).
Such crystalline silicates are considered to be formed at the innermost region of the disk, as they require temperatures above 800 K (Hallenbeck et al. 1998). 
The shock heating allows the crystallization of amorphous silicates locally at the outer regions of the disk.
{ The required velocity is indicated by a dotted line in Figure \ref{fig:introvrel} (see \S 4 for detail).
Even without the secular resonance, nearly half the planetesimals achieve the required velocity to contribute toward the formation of crystalline silicates.}
Tanaka et al. (2013) proposed the evaporation of planetesimals because of bow shocks associated with planetesimals orbiting with supersonic velocities relative to the gas in a protoplanetary disk.
{ The required velocity is indicated by a solid line in Fig. \ref{fig:introvrel}. 
The icy planetesimals perturbed by Jupiter easily acquire the required velocity.}
If the 100-km-sized icy planetesimals obtain a relative velocity of $\ga$2 ${\rm kms^{-1}}$ with respect to the gas disk, they evaporate in a timescale of $10^7$ yr even outside the snow line . 
Tanaka et al.(2013) revealed the possibility that active evaporation of planetesimals in the vicinity of the asteroid belt could change icy planetesimals with a core-mantle structure to rocky planetesimals and resolve the problem of overabundance of water in the terrestrial planets formed in the cold disk.

{ Although the supersonic planetesimals can cause these important effects in the primordial solar system,} previous studies of the shock-generating planetesimals were performed with the limited parameters of the current Jupiter.
Actually, Jovian planets have possibly undergone migration during their formation and the location where proto-Jupiter of our solar system was formed is not well known.
If we can constrain the orbital parameters of proto-Jupiter, which meets the conditions required for generating chondrules, we could obtain important information about the early stage of Jovian formation.
It is interesting to know the conditions in which evaporation occurs in extrasolar systems. 
The evaporation of icy planetesimals would release molecules, which do not normally exist in the gas phase. 
If sufficient amounts of icy planetesimals evaporate from a suitable location, the molecules known as shock tracers would be detectable through Atacama Large Millimeter/submillimeter Array (ALMA) observations. 
To estimate the possibility of observation, we should know the total amount of evaporation, the location where the planetesimals have a large velocity, and the length of the period of depletion.
Unfortunately, previous studies on the evolution of planetesimals perturbed by Jovian planet only targeted the solar system parameters. 
As the location of a Jovian planet and the density of the gas disk control secular resonance, more detailed simulations of the planetesimals crossing the resonances of extrasolar Jovian planets with varying parameters and in different disk masses are required to compare the data from ALMA observations.
It is not known whether the asteroid belt is common in exosystems or is special to our solar system. 
Analogous to our solar system, the formation of extrasolar Jovian planets affects the population and dynamical distribution of exo-asteroids. 
If we can detect the evaporated ice materials, chemically evolved molecules, collisional dust rings, or heat released by the collisions together with proto-Jupiter in the exosystems, it could help determine the history of formation of the asteroid belt.
To connect the observations of exosystems to our asteroids, studies on planetesimal evolutions caused by the extrasolar Jovian planets with various masses, eccentricities, and semi-major axes are necessary.

In this study, we numerically examined the orbital evolutions of planetesimals in the asteroid region under the influences of the Jovian planet, protoplanetary disk gravity, and gas drag.
In Nagasawa et al. (2014), we only considered a Jovian planet with current size and orbital parameters. In addition, the planetesimal size was limited to 300 km and all planetesimals were located at 4.1 au. 
The gas drag formula they used is not suitable for high-speed planetesimals.
In this paper, although we have considered our solar system as the base, we have changed the density of the gas disk, the planetesimal size and location, and mass, eccentricity, and semi-major axis of the Jovian planet. 
We also adopt the gas drag formula that depends on the Mach speed.
The eccentricity evolution of asteroids is controlled by excitation because of the resonances and damping due to the gas drag. 
In section \ref{sec:basiceqs}, we describe the equations used in the numerical simulations. 
The numerical results are shown in section \ref{sec:results}.
Secular resonance excites the asteroidal eccentricity up to $e\sim$0.6--0.8 according to our solar system parameters for planetesimals larger than 100 km. 
The maximum eccentricity is normally achieved at the location of 3:1 resonance.
The maximum speed of the planetesimal in the gas disk depends on the initial planetesimal parameters, the disk density, and the mass, eccentricity, and semi-major axis of the Jovian planet. 
In section \ref{subsec:astparam}-\ref{subsec:jupiter}, we describe the effect of varying the model parameters.  
{ Bearing in mind these parameter analyses, we return to the conditions for chondrule formation and planetesimal evaporation in section \ref{sec:discussions}}.
Finally, we summarize the results in section \ref{sec:conclusions}.

\vspace{2mm}

\section{BASIC EQUATIONS}
\label{sec:basiceqs}

We investigated the orbital evolution of a planetesimal (10--1000 km) perturbed by a Jovian planet and the protoplanetary disk neglecting the relativity. 
We treated the planetesimal as a test particle. 
The forces per unit mass in the equations of motion for planetesimal $i$ and Jovian planet p in the heliocentric coordinate are obtained as
\begin{eqnarray}
&&{\bf F}_i= -\frac{G(M_*+m_i)}{r_i^3}{\bf r}_i
  +GM_{\rm p} \frac{{\bf r}_{\rm p}-{\bf r}_i}{|{\bf r}_{\rm p}-{\bf r}_i|^3} 
  -GM_{\rm} \frac{{\bf r}_{\rm p}}{|{\bf r}_{\rm p}|^3}
+{\bf F}_{{\rm disk},i}+{\bf F}_{{\rm drag},i}
\label{eq:equmoi}, \\
&&{\bf F}_{\rm p}= -\frac{G(M_*+m_{J})}{r_{\rm p}^3}{\bf r}_{\rm p} +{\bf F}_{{\rm drag, p}}.
\label{eq:equmoJ}
\end{eqnarray}
The first term in Equation (\ref{eq:equmoi}) is the gravity from the star, the second term is the gravity from a Jovian planet, and the third term is an indirect term that appears from the motion of the Jovian planet.
The term ${\bf F}_{\rm disk}$ is the force caused by self-gravity of the disk and ${\bf F}_{\rm drag}$ is the gas drag.

We consider an axially symmetric disk whose origin is placed at the central star. 
Its surface density is $\Sigma=\Sigma_1 (R/{\rm 1 au})^{-\alpha}$ and the temperature is $T=T_1 (R/{\rm 1 au})^{-\beta}$ , where $R$ denotes the distance from the z-axis at the cylindrical coordinate. 
The disk rotates with an angular velocity of $\Omega$. 
The force caused by the self-gravity of the disk, ${\bf F}_{\rm disk}$, is given as (Ward 1981)
\begin{equation}
{\bf F}_{{\rm disk}}=-4\pi G Z_{\alpha}\Sigma {\bf e}_{R},
\label{eq:diskgrav}
\end{equation}
where $G$ is the universal gravitation, ${\bf e}_{R}$ is a unit vector in the  $R$ direction, $Z_{\alpha}$ takes the values of $Z_{1}=1$, $Z_{2}=2$, $Z_{1/2}=Z_{5/2}=0.685$, and $Z_{3/2}=1.094$ for $\alpha=1$, 2, 1/2, 5/2, and 3/2, respectively. 
We used $\alpha=3/2$, $\beta=1/2$, and $T_1=280$ K in our simulations. 
We used $\Sigma_1=\Sigma_{\rm MMSN}$ unless otherwise noted, where $\Sigma_{\rm MMSN}$ is the surface density of the minimum mass disk at 1 au, $1700$ $\rm {gcm^{-2}}$. 
The location of secular resonance, which excites the eccentricity of planetesimals, insensitively depends on the disk density and density profile.
Secular resonance occurs between 2 and 4 au in a wide range of the disk profile (Nagasawa, Tanaka, \& Ida 2000; Nagasawa, Lin, \& Thommes 2005). 
Even when we choose a different power for the disk density, its effect on the maximum velocity of the planetesimal would be limited.
A gap in the disk formed by the Jovian planet is ignored for simplicity because the peak velocities of the planetesimals are generally achieved far from a Jovian planet  (Fig. \ref{fig:reso}, Marzari \& Weidenschilling 2002; Nagasawa et al. 2014) and the gap (Kanagawa et al. 2017).  
The protoplanetary disk provides a background gravitational field and gas drag. 
{ Actually, the density of the disk near the planet would not be uniform, unlike our assumption here. Although the precession rate of the pericenter and strength of the gas drag would change occasionally in the disk structures, the timescales of secular evolution and the migration due to gas drag are normally much longer than the timescales in which the planetesimals pass the structure.
We suppose that the averaged density profile of the orbit determines the evolution.
Even if the gap is formed, secular resonance occurs at similar locations (e.g., Ward 1981; Nagasawa, Tanaka, \& Ida 2000). 
However, the planetesimals in the gap do not receive the gas drag until their eccentricities are enhanced and they move beyond the gap.}

Although we calculated the orbits of bodies in three dimensions, we used the gravity of a thin disk, i.e., we considered the z-component of force caused by the gravity of the disk as negligible. 
The thin disk approximation is reasonable in simulations that do not include disk gravity, as the bodies maintain almost coplanar orbits. 
In simulations that include the disk's self-gravity, the planetesimals temporarily go up $\la 20^{\circ}$ above the disk's scale height.
In this study, we focused on the eccentricity evolutions and not the vertical excursion by using the thin disk approximation.
We did not dissipate the disk during individual simulation; instead, we performed simulations by using different masses of the disk and locations of the planetesimals as the initial conditions.
This is because the time-scale of {\it sweeping} secular resonance is of the order of the gas depletion time-scale ($\sim 10^6-10^7$ yr) and is more than the orbital decay time of the planetesimals ($\sim 10^4-10^6$ yr). 

The velocity difference between the gas disk and celestial body results in a gas drag.
By using relative velocity ${\bf v}_{\rm rel}={\bf v}-{\bf v}_{\rm gas}$, the gas drag force was given by Adachi et al. (1976) as
\begin{equation}
{\bf F}_{\rm drag}=-\frac{1}{2}C_{\rm D}\pi \rho r_{\rm a}^2 m^{-1}v_{\rm rel}{\bf v}_{\rm rel},
\label{eq:dragforce}
\end{equation}
where $m$ is the planetesimal mass, $r_{\rm a}$ is the planetesimal radius, $\rho$ is the density of the gas disk (Equation (\ref{eq:diskdensity})), $C_{\rm D}$ is the drag coefficient, and $ v_{\rm rel}$ is the magnitude of ${\bf v}_{\rm rel}$. The value of $C_{\rm D}$ is calculated as
\begin{equation}
C_{\rm D}= \left[ \left( \frac{24}{R_e}+\frac{40}{10+R_e}\right)^{-1}+0.23 M_a \right]^{-2}+\frac{2.0 (0.8 \kappa+M_a)}{1.6+M_a},
\end{equation}
where $\kappa$ is the correction term with a value of 0.4 for the Reynolds number $R_e<2 \times 10^5 $ and 0.2 for $R_e>2 \times 10^5 $, 
$R_e=(\pi/2)^{1/2}0.353^{-1} r_{\rm a} M_a l_{\rm g}^{-1}$, $l_{\rm g}$ is the mean free path, the Mach number $M_a= v_{\rm rel}/c_{\rm s}$, and $c_{\rm s}$ is the isothermal speed of sound (Tanigawa et al. 2014).
Nagasawa et al. (2014) used a weaker gas drag than this formula, i.e., their 300-km-sized planetesimals correspond to larger planetesimals in this paper. 
The density of a planetesimal is $ \rho_{\rm mat}=2\ {\rm g cm^{-3}}$. 
The material density was used only in the evaluation of the radii of bodies from their masses. 

The gas velocity is estimated as follows.
By using the balance of the centrifugal force, central star gravity, disk gravity, and pressure gradient, we can obtain equations in the $R$ and $z$ directions as
\begin{eqnarray}
&&\Omega^2R-\frac{GM_*}{r^2}\frac{R}{r}-\frac{1}{\rho}\frac{\partial P}{\partial R}+F_{{\rm disk},R}=0,
\label{eq:diskR}\\
&&-\frac{GM_*}{r^2}\frac{z}{r}-\frac{1}{\rho}\frac{\partial P}{\partial z}=0,
\label{eq:diskz}
\end{eqnarray}
where $R$ is related to $r$ by $r^2=R^2+z^2$, $F_{{\rm disk},R}$ is the $R$ component of the disk gravity,  $P$ is the pressure, $M_*$ is the mass of the central star, and $\Omega$ is the angular velocity.

We considered a sufficiently large disk, in which the sound velocity $c_{\rm s}$ and temperature $T$ are independent of $z$. 
By using $c_{\rm s}^2=kT/\mu m_{\rm u}$, $P=c_{\rm s}^2 \rho$, the Boltzmann constant $k$, mean molecular weight $\mu=2.24$, atomic mass unit $m_{\rm u}$, scale height $H=\sqrt{2} c_{\rm s}/ \Omega_R$, and $\Omega_R=(GM_*/R^3)^{1/2}$,
 the partial derivative of $P$ with respect to $R$ is obtained from equation (\ref{eq:diskz}) as
\begin{equation}
\frac{1}{\rho R}\frac{\partial P}{\partial R}=-\left( \frac{C_{\rm s}}{R}\right)^2 \left( \alpha +\frac{3}{2} +\frac{\beta}{2}\right)-\left(\frac{\beta}{2}-\frac{3}{2}\right)\left( \frac{\Omega_{K}z}{R}\right)^2.
\label{eq:raundpr}
\end{equation}
Here, we used $(GM_*/R^3)^{1/2} \sim \Omega_K \equiv (GM_*/r^3)^{1/2}$ in the pressure gradient term, as the term is small compared with the stellar gravity.
The disk density is given as
\begin{equation}
\rho=\frac{\Sigma_1}{\sqrt{\pi}H}\left(\frac{R}
{\rm 1 au}\right)^{-\alpha}\exp\left(-\frac{z^2}{H^2}\right),
\label{eq:diskdensity}
\end{equation}
where $\Sigma_1$ is the disk surface density at 1 au.
For the MMSN disk, $\Sigma_1$ is equal to 
$\Sigma_{\rm MMAN} = 1700$ $\rm {gcm^{-2}}$ (Hayashi et al. 1985).
In addition, Equation (\ref{eq:diskR}) gives
\begin{equation}
\Omega^2=\Omega_{K}^2\left[1-\left( \frac{\beta}{2}-\frac{3}{2} \right)
\left(\frac{z}{R} \right)^2
-\left( \frac{c_{\rm s}/R}{\Omega_K}\right)^2 \left( \alpha +\frac{3}{2} +\frac{\beta}{2}\right)+\frac{F_{{\rm disk},R}}{\Omega_K^2 R}
 \right].
\label{eq:omega2}
\end{equation}
Equation (\ref{eq:omega2}) is equivalent to the equation shown by Tanaka et al. (2002) when $F_{{\rm disk},R}=0$. The second and later terms in the square brackets of Equation (\ref{eq:omega2}) are small quantities. The rotation speed of the disk can be rewritten as
\begin{equation}
\Omega \simeq \Omega_K \left[ 1-(\eta_{\rm p}+\eta_{\rm d}) \right],
\end{equation}
where
\begin{eqnarray}
&& \eta_{\rm p} \equiv \frac{1}{4}(\beta-3)\left( \frac{z}{R}\right)^2
          +\frac{1}{4}(2\alpha+3+\beta)\left( \frac{c_s}{R\Omega_R}\right)^2,\\
&& \eta_{\rm d} \equiv \frac{F_{{\rm disk},R}}{2\Omega_K^2R}.
\end{eqnarray}

\vspace{2mm}

\section{NUMERICAL RESULTS}
\label{sec:results}

\begin{figure}\epsscale{0.8}
\plotone{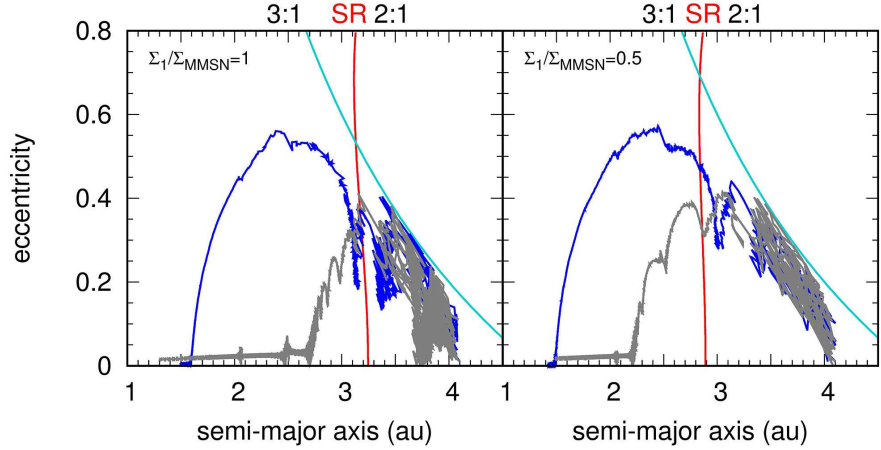}
\caption{Evolution of $e$ versus $a$ for a 100-km planetesimal originating at $a=4.1$ au. 
Left: The density of the minimum-mass disk is assumed ($\Sigma_1=\Sigma_{\rm MMSN}$). Right: The disk density is 50\% of the minimum mass disk ($\Sigma_1=0.5 \Sigma_{\rm MMSN}$).
In both panels, the blue curve shows the evolutions of the planetesimals, in which disk gravity is considered, and the gray curve shows the simulations in which the disk gravity is neglected (the gas drag is included).
Each simulation time is $3 \times 10^7$y. Locations of secular resonance (red line) and 2:1 and 3:1 mean-motion resonances (dotted lines) are also shown. 
{ The light-blue line represents $a(1+e)=$4.8 au, which shows crossing of the Jovian Hill's radius.}
(A colored version of this figure is available in the online journal.)
\label{fig:wwodisk}}
\end{figure}

The orbits of the planetesimal and Jovian planet were integrated with the 4th-order Hermite code. 
The mass of the central star was fixed at 1 solar mass (1 $M_{\rm \odot}$). 
Although the stellar mass was a fixed value, our results can easily be scaled or extended to consider other cases because the locations of secular and mean-motion resonances can be scaled in a classical manner.
Hereafter, planetesimals are initially set at 4.1 au, unless otherwise stated; 
this is just beyond 3:2 resonance.
We express the eccentricity, semi-major axis, and mass of the Jovian planet as $e_{\rm p}$, $a_{\rm p}$, and $M_{\rm p}$, respectively. The planetary mass is measured by using Jovian mass $M_{\rm J}$.
Figure \ref{fig:wwodisk} shows the typical evolution of 100-km-sized planetesimals under the influence of a Jovian planet with $e_{\rm p}=0.048$, $M_{\rm p}=1$ $M_{\rm J}$, and the gas disk. 
The evolutions of planetesimals with and without the disk's self-gravity are shown by gray and blue curves, respectively. 
The disk densities differ between the left and right panels.
In the left panel, the disk density is that of the minimum-mass solar nebula disk and secular resonance of the Jovian planet occurs near 3.2 au (red vertical line).

The planetesimal migrates inward due to gas drag from 4.1 au, passing through several mean-motion resonances existing between 3 and 4 au. 
These resonances increase the eccentricities of the planetesimals. 
As the separation of the resonances increases with the increase in distance between the planetesimals and the Jovian planet, the eccentricity damping due to gas drag exceeds its excitation because of the mean-motion resonances at some points.

Although they did not take into account the effect of secular resonance (i.e., disk gravity), the detailed works by Marzari \& Weidenschilling (2002) revealed the process by which the asteroids gain high eccentricity by passing through resonance and resonance trapping (gray curve).
The 100--300-km-sized planetesimals that reach 2:1 resonance with an eccentricity of $\ga$ 0.2 can gain additional eccentricity and attain values of eccentricity ($e$) larger than 0.4 at 2:1 resonance. 
The 2:1 trapping requires a noncircular Jovian orbit ($e_{\rm p} \ga 0.03$) and slow drifting of the planetesimals ($\ga$ 100 km).
The excitations of eccentricity before 2:1 resonance, which are caused by the crossing of multiple resonances, are insensitive to the planetesimal sizes. 
However, the evolution in 2:1 resonance depends on the planetesimal size.
The trapping time in 2:1 resonance is proportional to the planetesimal diameter. 
A longer 2:1 trapping time results in higher eccentricity. 
In our simulation (Fig. \ref{fig:wwodisk}) using the drag force expressed by Equation (\ref{eq:dragforce}), the planetesimal eccentricity rises up to 0.2--0.4 during libration in 2:1 resonance, and the planetesimals eventually escape from the resonance and are circularized in the region between 2 and 3 au owing to the gas drag. 
Their maximum eccentricities are in the range of 0.3--0.5.

If the effect of secular resonance is considered, the maximum velocities of the planetesimals are higher (blue curve) than simulations without disk gravity (gray curve). 
Secular resonance moves planetesimals from 2:1 to 3:1 resonance while retaining the high eccentricity.
As 3:1 resonance at $\sim$ 2.5 au is far from the other resonances, the eccentricity is damped before the planetesimal reaches 3:1 resonance unless we include the self-gravity of the disk.
However, when the disk's gravity is considered, secular resonance occurs and the planetesimals evolve along the line of $a(1+e) \sim 4.8$ au until near 3:1 resonance. 
The light-blue line in Figure \ref{fig:wwodisk} represents the location where an apocenter of the planetesimal is at $a(1+e)=4.8$ au, which shows that the crossing is beyond the Jovian Hill's radius.
When the disk gravity is considered, 90\% planetesimals in our 60 simulations record the maximum eccentricity inside the 5:2 resonance ($\sim$2.8 au) with $e_{\rm max}\sim$0.4--0.6, whereas this is reduced to 20\% when the disk gravity is neglected. 
Secular resonance enables the planetesimal to gain higher relative velocity with respect to the surrounding gases. 
The difference becomes clearer in larger planetesimals.
In the case of 300-km-sized planetesimals under the influence of secular resonance, 92\% and 75\% planetesimals record the maximum eccentricity inside the 5:2 resonance ($\sim$2.8 au) and 3:1 resonance ($\sim$2.5 au), respectively. The planetesimals have $e=0.62$ on average and they are distributed from $e_{\rm max}\sim 0.5$ to 0.7. Only a few percent of planetesimals can reach 3:1 resonance without secular resonance. In this case, $e_{\rm max}\sim 0.3-0.5$, and the average value of $e_{\rm max}$ is 0.45.

The inclination here is slightly enhanced around 3 au when we include the disk gravity. 
This enhancement is $\la 10^{\circ}$ unless a strong encounter with a Jovian planet occurs. These inclinations are smoothly damped by the gas drag near the terrestrial region.
When planetesimals have non-negligible inclination, the disk density is low near the apocenter or pericenter of the planetesimal. 
{ The energy flux of planetesimal bow shocks is proportional to the gas density, $\rho$, as $\rho v_{\rm rel}^3$.
The required relative velocities to melt chondrule precursors and evaporate icy planetesimals are proportional to $\rho^{-1/5}$ (Iida et al. 2001) and $\rho^{-1/3}$ (Tanaka et al. 2013), respectively. 
When the orbits are inclined , there is a possibility that sufficient shock is not generated in all periods, as the planetesimals move between dense (near mid-plane) and less dense areas (above) of the disk.}

However, the relative velocity is minimal at the apocenter and pericenter (\S 3.2).
The thin disk approximation that we used to calculate the disk gravity near the location where the planetesimals generate the shock waves is valid. 
Moreover, as the gas drag also weakens in the thin area, the durations of high eccentricities are lengthened when planetesimals have high inclinations. 

The right panel of the figure shows the case in which the disk density is half the minimum-mass solar nebula disk. 
Secular resonance occurs near 2.9 au in this case. 
With this disk density, the gas drag force is approximately half that of the case in the left panel. 
When the secular resonance is included, the maximum eccentricity ($e_{\rm max}=$0.45 on average) is reached around the 3:1 resonance (2.7 au on average). 
Without the secular resonance, the average of $e_{\rm max}$ is 0.39, which is reached at 3.2 au on average. 
The secular resonance that occurs inside 3 au can still move the planetesimals until the 3:1 resonance.
The maximum eccentricity does not sensitively depend on the disk mass (see Sec. \ref{subsec:diskrho}) as long as the self-gravity of the disk is included.
These panels show that secular resonance between 3:1 and 2:1 resonances results in higher relative velocity of the planetesimals.

\begin{figure}\epsscale{0.8}
\plotone{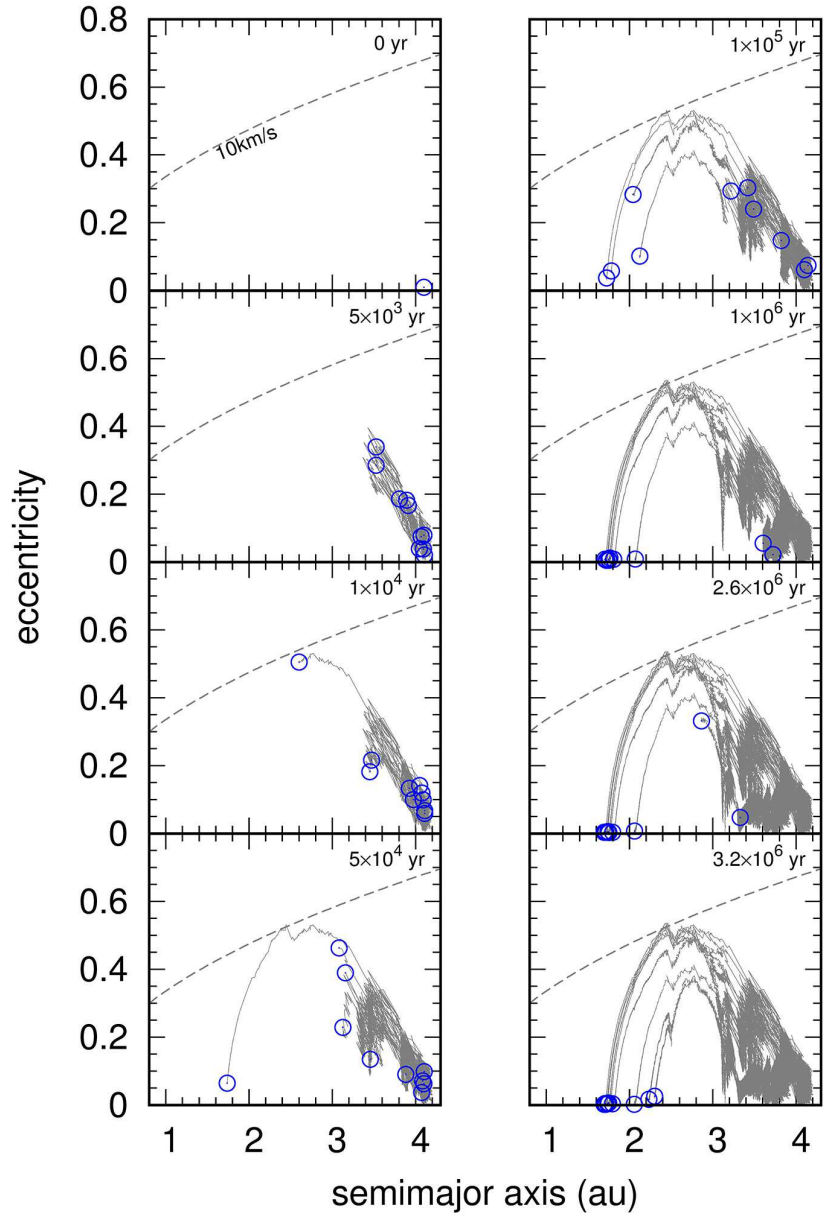}
\caption{Snapshot of the evolution of planetesimals in the $e$-$a$ plane. Ten planetesimals 100 km in size (blue circles) and originating at $a=4.1$ au are shown with their trajectories (gray curves). 
The dotted lines show the eccentricities whose relative velocities with respect to the gas ($e v_{\rm kep}$) equals 10 ${\rm km s^{-1}}$.
 (A colored version of this figure is available in the online journal.)
\label{fig:snap}}
\end{figure}

Owing to resonance trappings, the timing of high velocity differs from particle to particle, and the total time of migration is $\sim 10^6$ yr for planetesimals of 100-km size.
Figure \ref{fig:snap} shows the snapshots of 10 planetesimals of 100-km size in the $e$-$a$ plane.
The dashed line represents the location where $e v_{\rm kep}= e(GM_{\rm \odot}/a)^{1/2}= $10 ${\rm km s^{-1}}$.
Initially ($t=0$), the planetesimals are at $a=4.1$ au with $e=0.01$. 
The Jovian planet has the current orbital parameters of Jupiter, and a minimum-mass disk is assumed.
At approximately $t=10^4$ yr, one of the planetesimals reaches the highest eccentricity at 3:1 resonance with relative velocity of 10 ${\rm kms^{-1}}$.
Other planetesimals are trapped in 2:1, 3:2, or other resonances beyond 3 au.
At 1 Myr, eight of the ten planetesimals move to within $\sim$2 au, while the rest remain at approximately 4 au.
When the eccentricity is totally damped ($e \simeq 0$), the migration is stopped because there is no relative velocity with respect to the gas disk.
In this figure, all planetesimals finish migration after 3.2 Myr.
The planetesimals end their journey at the terrestrial region, and the location where a planetesimal stops depends on how highly the eccentricity is excited.
Higher excitation implies longer migration.
The duration for which an individual planetesimal (100 km) has high velocity is $\sim 10^4$ yr $- 10^5$ yr; however, the event is continued for $\sim 10^6$ yr because the timing of high velocity varies between planetesimals (see \S \ref{sec:discussions}).

\clearpage

\subsection{Mean-Motion and Secular Resonances}
\label{subsec:reso}

\begin{figure}\epsscale{0.8}
\plotone{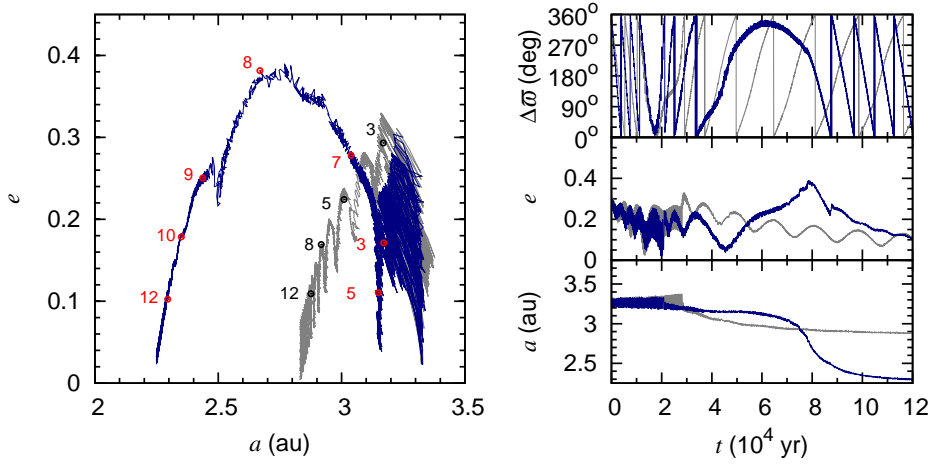}
\caption{Time evolution of orbital parameters. 
The evolutions with disk gravity and without disk gravity are shown by the blue curve and gray curve, respectively. 
Left: Evolution of $e$ versus $a$ for a 100-km planetesimal starting from 2:1 resonance. 
The numbers along the curves show the time ($\times 10^4$ yr) at the corresponding points. 
Top Right: The difference in longitude of the pericenter between the planetesimal and the Jovian planet. 
Middle Right: Eccentricity evolution.
Bottom Right: Semi-major axis evolution. 
 (A colored version of this figure is available in the online journal.)
\label{fig:periae}}
\end{figure}

Figure \ref{fig:periae} shows the time evolution of the longitude of the pericenter (top right panel), eccentricity (middle right panel), and semi-major axis (bottom right panel) of a 100-km-sized planetesimal.  
The simulation is initiated from the location of 2:1 resonance.
The evolution on the $e$-$a$ plane is shown in the left panel, where 
the numbers represent time in the unit of $10^4$ yr.
The cases with (blue curve) and without disk self-gravity (gray curve) are calculated with the same orbital parameters.
Secular resonance occurs when the rotation speeds of the pericenter of the Jovian planet and a planetesimal almost coincide (top right panel, $\sim 4 \times 10^4$ yr$ - 8 \times 10^ 4$ yr).
The eccentricity drastically changes during this period. 
Such a coincidence does not occur without the disk gravity. 
In this case, the relative rotation speed simply decreases according to the distance from the Jovian planet.

The trapping period in 2:1 resonance depends on various situations, as shown in Figure \ref{fig:snap}. 
In Figure \ref{fig:periae}, the planetesimal is trapped in 2:1 resonance until $3 \times 10^4$ yr when disk gravity is not included (the semi-major axis oscillates between 3.2 au and 3.3 au).
 The planetesimal is trapped in secular resonance for a while when the disk self-gravity is included. 
At $e \sim 0.3$, the gas drag increases to the point when the planetesimal migrates out of the resonance. 
The planetesimal has high eccentricity ($e>0.3$) in the interval of approximately $10^4$ yr.

\begin{figure}\epsscale{0.7}
\plotone{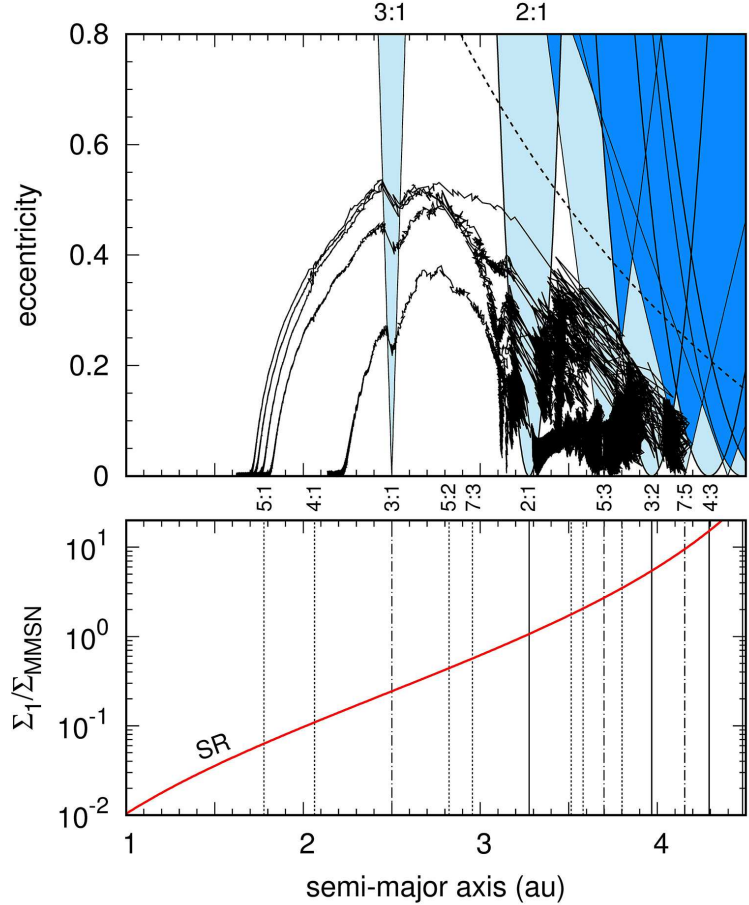}
\caption{Location of the mean-motion resonances and $\nu_5$ secular resonance together with five orbital calculations. 
The light-blue areas show the widths of the mean-motion resonances. 
The blue areas show the overlapping areas of the mean-motion resonances. 
The dotted line in the top panel shows the parameters where the apocenters of the planetesimals reach the Jovian semi-major axis (5.2 au). 
In the bottom panel, the first-order mean-motion resonances ($j+1:j$), second-order resonances ($j+2:j$), and others are shown by solid lines, dotted-dashed lines, and dotted lines, respectively. 
The vertical axis shows the ratio of the density to the minimum mass of the solar nebula model. 
The red thick solid line in the bottom panel shows the location of $\nu_5$ resonance. 
 (A colored version of this figure is available in the online journal.)
\label{fig:reso}}
\end{figure}

Although the exact location of the mean-motion resonance rarely depends on the disk mass, the location of secular resonance strongly depends on the disk mass. 
We show the locations of $\nu_5$ secular resonance and mean motion resonances caused by current Jupiter in Figure \ref{fig:reso}. 
The top panel shows the widths of the mean-motion resonances, which are obtained through simple pendulum equations by neglecting the contributions from the mean longitude at the epoch and longitude of the pericenter (Murray \& Dermott 1999). 
The exact locations of the mean-motion resonances are separately shown in the bottom figure. 
The first-order mean-motion resonances ($j+1:j$), second-order resonances ($j+2:j$), and other resonances are represented using solid, dotted-dashed, and dotted lines, respectively. 
The vertical axis of the bottom figure represents the ratio of the density to the minimum mass of the disk (minimum mass of the solar nebula model). 
In the case of minimum mass of the disk, $\nu_5$ exists at 3.3 au. 
The planetesimals, which initiate migration from $\ga $ 4.1 au, can experience secular resonance even in a disk that is approximately 10 times massive than the minimum-mass disk. 
When the disk density is between 20\% and 100\% of the minimum-mass disk, planetesimals can experience secular resonance between 3:1 and 2:1 resonances.

Inside the mean-motion resonance, the semi-major axis and eccentricity librate and evolve upward to the left (top panel), which is the same as the evolution shown by Marzari \& Weidenschilling (2002). 
The libration in the mean-motion resonance keeps the planetesimal's apocenter away from the Jovian planet as it retains the Tisserand parameter. 
The dotted line in the upper panel of Figure \ref{fig:reso} shows the location at which the planetesimal's apocenter reaches 5.2 au (orbit crossing with the planet).
The overlapping area of the major mean-motion resonance is represented using a darker blue color. 
In actuality, the planetesimals avoid the major overlapping area.

\begin{figure}\epsscale{0.7}
\plotone{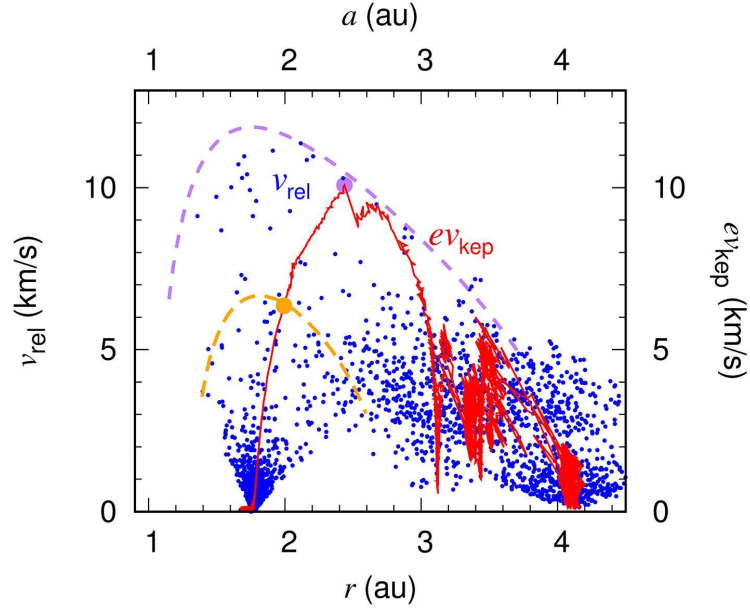}
\caption{Example of relative velocity with respect to the gas disk.
The blue dots show the relative velocity in approximately every 200 yr obtained from a single run ($r$ vs $v_{\rm rel}$).
The red curve shows the same evolution in the $a$ vs $ev_{\rm kep}$ plane.
The purple and yellow circles show the points of $e\sim 0.53$ and $e \sim 0.3$, respectively.
With this $e$ and $a$, the relative velocity varies along the same colored dashed curves during one orbit.
(A colored version of this figure is available in the online journal.)
\label{fig:vrelecc}}
\end{figure}

\subsection{Relative Velocity and Eccentricity}
\label{subsec:velocity}

In this paper, we mainly focus on eccentricity to consider the relative velocity.
In reality, the absolute value of the velocity of the planetesimal relative to the gas disk ($v_{\rm rel}$) changes during orbit from approximately $1/2 e v_{\rm kep}$ to $e v_{\rm kep}$ in the case of planetesimals orbiting in the mid-plane of the disk. The Keplar velocity is evaluated using the semi-major axis as $v_{\rm kep}=(GM_{\odot}/a)^{1/2}$.
The planetesimal in the mid-plane rotates faster than the gas disk at the pericenter, and rotates slower than the disk at the apocenter.
In the lowest order of $e$, $v_{\rm rel}$ is approximately $1/2 e v_{\rm kep}$ at both the pericenter and apocenter; this is the minimum value.
The maximum velocity, $v_{\rm rel} = e(1-e^2)^{-1/2} v_{\rm kep} \sim e v_{\rm kep}$, is achieved at $e=\cos{u}$, where $u$ is the eccentric anomaly. 
{ When the eccentric anomaly is specified as $\cos{u}=e$, the distance is $r=a(1-e^2)$.}
Although a typical relative velocity is often referred to as $ e v_{\rm kep}$, the averaged relative velocity is $v_{\rm rel} \sim 0.77 e v_{\rm kep}$. 
If we evaluate the Kepler velocity using the radial distance $r$ instead of $a$, for example, $V_{\rm kep, r}=(GM_{\odot}/r)^{1/2}$, the maximum relative velocity is exactly $eV_{\rm kep, r}$.
The maximum relative velocity is $e v_{\rm kep}$ in the lowest order of $e$; however, it becomes underestimated with rise in eccentricity. 
For example, if $e=0.6$, $e(1-e^2)^{-1/2} v_{\rm kep}=0.75 v_{\rm kep}$ whereas $e v_{\rm kep}=0.6 v_{\rm kep}$.

Figure \ref{fig:vrelecc} illustrates the simulation in two ways for $v_{\rm rel}$-$r$ and $e v_{\rm kep}$-$a$. 
The initial inclination of the planetesimal is $i=0.01$ rad. 
Although we include the $z$-component of velocity in the calculation of $v_{\rm rel}$, the relative velocity caused by the inclination is much smaller than that caused by the eccentricity.
The left axis shows the values of $v_{\rm rel}$ (blue dots) obtained by a typical run at a distance of $r$ from the star, which is represented in the bottom axis. 
The right axis represents $e v_{\rm kep}$ (red curve) at the semi-major axis shown by the top axis.
Blue dots ($v_{\rm rel}$-$r$) are plotted every 200 yr in one simulation shown by the red curve ($e v_{\rm kep}$-$a$).
In this simulation, when the planetesimal has the maximum eccentricity, its semi-major axis is $a\sim 2.4$ au and eccentricity is $e\sim 0.53$.
In the $a$-$e v_{\rm kep}$ plane, this position is shown by a solid purple circle at $e v_{\rm kep} \sim 10$ ${\rm kms^{-1}}$ at 2.4 au (the peak of the red curve). 
The planetesimal evolves along the purple dashed curve in the $r$-$v_{\rm rel}$ plane with this semi-major axis and eccentricity, i.e., during the orbit, the planetesimal travels from $r=a(1-e)\sim 1.1$ au to $r=a(1+e)\sim 3.7$ au along the purple curve. 
The top relative velocity is $v_{\rm rel} = e(1-e^2)^{-1/2} v_{\rm kep} \sim 12$ ${\rm kms^{-1}}$ at $r=a(1-e^2) \sim 1.8$ au.
Thus, all blue dots in the terrestrial region are below the purple dashed curve. 
The velocity $e v_{\rm kep}$ is a {\it typical} relative velocity, but it does not represent the maximum velocity or the velocity on the purple dashed curve.
With a smaller eccentricity, $e v_{\rm kep}$ is close to the maximum velocity (solid yellow circle and yellow dashed curve).
Please note that the planetesimals specified by one set of values of $a$ and $e v_{\rm kep}$ actually move in a wide range with varying velocities.

\subsection{Dependence on planetesimal parameters}
\label{subsec:astparam}

The velocity evolution of the planetesimal depends on the initial conditions of the planetesimal, planet, and disk.
In this subsection, we show the dependence of velocity on the planetesimal size and the initial semi-major axis. 
We consider the Jovian planet as having the current orbital parameters of Jupiter.
As the size of the planetesimal increases, the gas drag experienced by it is weakened. 
Then, the migration slows and the trapping at 2:1 resonance tends to be longer. 
Although the values of maximum eccentricity depend on whether the planetesimals enter secular resonance, the aforementioned tendencies make it possible to obtain higher eccentricities, as shown by Marzari \& Weidenschilling (2002). 
As for the initial semi-major axis, the planetesimals that originate near the Jovian planet ($\ga $4 au) can obtain large eccentricities, whereas those originating inside 2:1 and secular resonance ($\sim $3 au) cannot obtain high eccentricities. 

\begin{figure}\epsscale{0.7}
\plotone{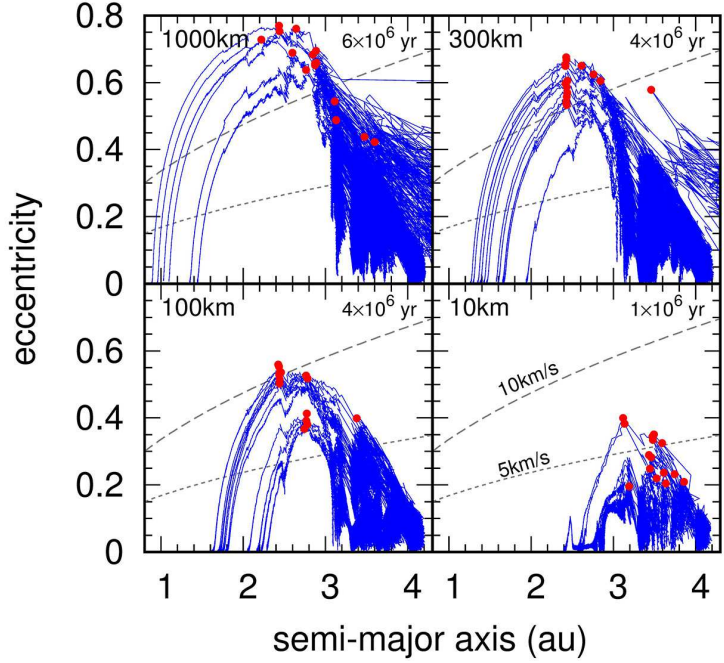}
\caption{Evolutions of 15 planetesimals with different sizes. 
The size of the planetesimals and the simulation time are indicated in each panel. 
The dotted lines and dashed lines are where the typical velocity of the planetesimals relative to the gas ($e v_{\rm kep}$) equals 5 ${\rm km s^{-1}}$ and 10 ${\rm km s^{-1}}$, respectively.
The maximum eccentricity of each simulation is noted by a red solid circle.
The initial semi-major axes of the planetesimals are 4.1 au.
 (A colored version of this figure is available in the online journal.)
\label{fig:size}}
\end{figure}

When considering solar system parameters (i.e., the minimum mass of the solar nebula and the current orbital parameters and mass of Jupiter), the icy planetesimals around 3:1 resonance with sizes between few tens of kilometers and several hundred kilometers gain high eccentricities. 
Figure \ref{fig:size} shows the evolutions of planetesimals with different radii ($r_{\rm a}=$ 10, 100, 300, and 1000 km). 
The initial semi-major axes are all 4.1 au. 
Further, two lines indicating $ev_{\rm kep} \sim 5$ ${\rm kms^{-1}}$ and 10 ${\rm kms^{-1}}$ are shown for reference. 
We performed 15 simulations for each case.
The solid red circles show the maximum eccentricities of each planetesimal recorded inside the Jovian orbit. 

The larger planetesimals are more excited, as the gas drag on them is less effective. 
There is a distinction between planetesimals with $r_{\rm a}=10$ km and $r_{\rm a}>100$ km.
That is, the gas drag efficiently damps the eccentricity in 10 km-planetesimals, and even secular resonance cannot move the planetesimals with high eccentricity to 3:1 resonance from 2:1 resonance.
In such cases, the maximum eccentricity is maintained at $\la 0.4$, and the eccentricity is completely damped and migration is ended at around the location of the main belt. 
The radius of 100 km is the transitional size.
About half of the 100 km-planetesimals attain $v_{\rm rel} \sim 10$ ${\rm kms^{-1}}$ and the migrations end at approximately 1.6 au.
Although the maximum eccentricity is achieved at 3:1 resonance similar to the case of 300-km and 1000-km planetesimals, the eccentricity at approximately 2.5 au is slightly smaller for the 100-km planetesimals owing to the stronger gas drag.
At the size of 300 km, almost all planetesimals attain $v_{\rm rel} \sim 10$ ${\rm kms^{-1}}$.
The larger planetesimal migrates for a longer distance until the circularization.

The planetesimals often remain longer at resonances beyond 4 au while maintaining $e \la 0.1$. 
However, once the planetesimals start the journey to higher eccentricity ($e \ga 0.5$), their migration speeds are determined based on the strength of the gas drag. 
Thus, the migration slows considerably when the eccentricity is damped inside 2:1 resonance. 
Figure \ref{fig:size} shows the orbital integrations until the eccentricity excitation by resonance is almost depressed. 
The total simulation time is shown inside the panels.

\begin{figure}\epsscale{0.6}
\plotone{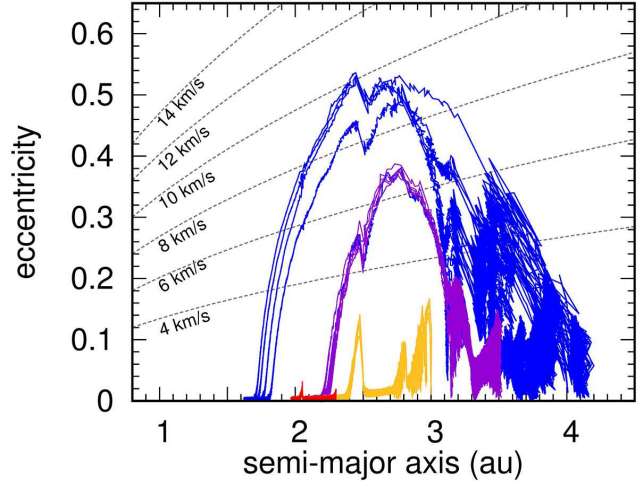}
\caption{ 
The orbital evolutions of planetesimals with different initial semi-major axes. 
The initial semi-major axes are 4.1 au (blue), 3.5 au (purple), 3.0 au (yellow), and 2.3 au (red). 
Five planetesimals (100 km) with different orbital angles are calculated in each case.
For illustration purpose, simulations are performed for $10^8$ yr at 2.3 au and 3.0 au and for $3 \times 10^7$ yr for other initial locations.
The planetesimals that originated within 3 au are not excited and their migrations take longer. 
The dotted lines show the typical relative velocities with respect to the gas ($e v_{\rm kep}$).
(A colored version of this figure is available in the online journal.)
\label{fig:aini}}
\end{figure}

\begin{figure}\epsscale{0.9}
\plotone{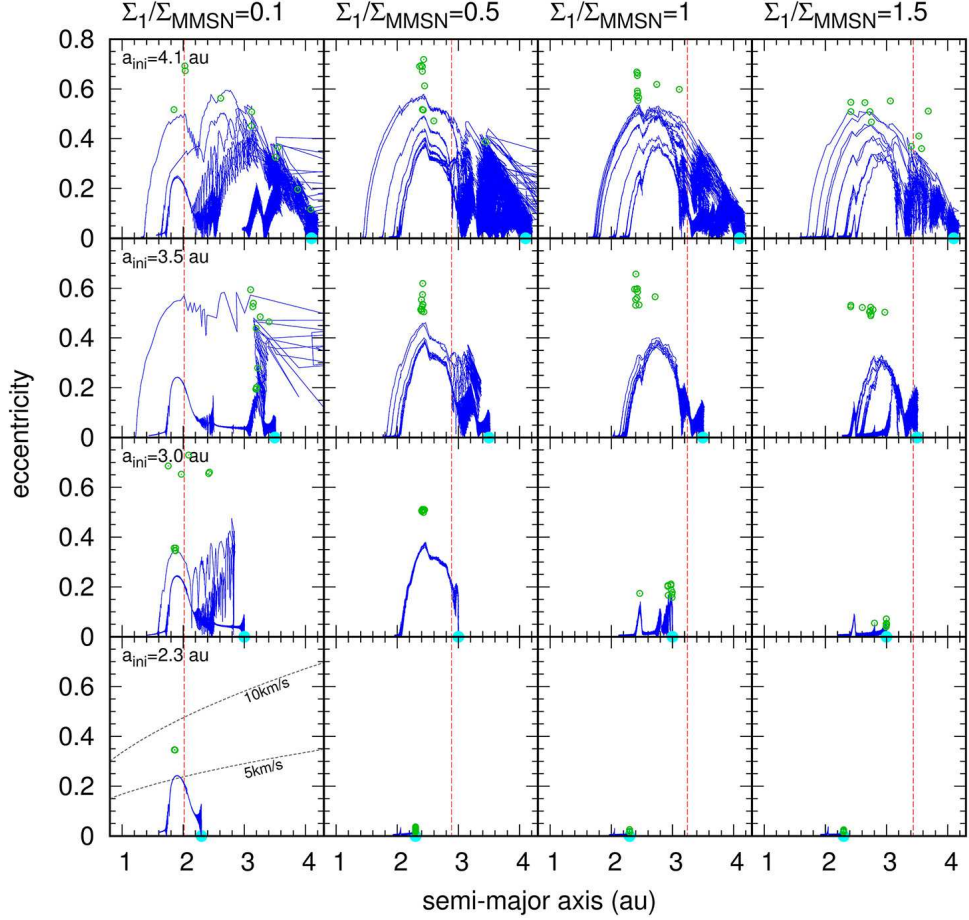}
\caption{ 
Trajectories of ten 100-km planetesimals with various disk densities and initial semi-major axes. 
From the left panel to the right panel, the disk surface densities are 0.1, 0.5, 1, and 1.5 times that of the minimum-mass disk. 
From the top panel to the bottom panel, the initial semi-major axes are 4.1 au, 3.5 au, 3 au, and 2.3 au.
The initial semi-major axes are shown by light blue solid circles in each panel.
The red dashed lines show the locations of secular resonance. 
The green circles show the maximum eccentricities of the ten 300-km planetesimals versus the semi-major axis where the maximum eccentricity is recorded. 
The 300-km planetesimals are excited more than the 100-km planetesimals.
(A colored version of this figure is available in the online journal.)
\label{fig:rho100}}
\end{figure}

When the planetesimals have different initial semi-major axes, their evolution paths differ because the resonances that the planetesimals pass through are different (see Figs. \ref{fig:aini} and \ref{fig:rho100}). 
The key resonances are $\nu_5$ secular resonance and 2:1 mean-motion resonance. 
The planetesimals inside both resonances are hardly excited (yellow and red curves in Fig. \ref{fig:aini}). 
The eccentricity grows most in the case where the planetesimal encounters multiple resonances near the Jovian planet, 2:1 resonance, and secular resonance (blue curves in Fig. \ref{fig:aini}).
This is because the eccentricity of a planetesimal that is excited before it crosses 2:1 resonance receives an additional boost (Marzari \& Weidenschilling 2002). 
The crossing of secular resonance is the most important factor for high excitation.
If the planetesimals pass through secular resonance, their passage through 2:1 or other resonances is not necessary for eccentricity excitation as long as the Jovian planet has the current eccentricity of Jupiter (purple curves in Fig. \ref{fig:aini}).
In contrast, when secular resonance cannot facilitate the excitation, the crossing of 3:2 followed by 2:1 resonances is required for the excitation of the eccentricity to more than 0.4, i.e., only the planetesimals beyond 4 au (0.8 $a_{\rm p}$) can obtain high eccentricity for the current Jupiter. 
Otherwise, the eccentricity remains less than $\sim$0.2 (Ida \& Lin 1996; Marzari \& Weidenschilling 2002). 

The planetesimals originating near the Jovian planet pass through many resonances and are highly excited. As a result, they migrate to a longer distance. 
In the area between 3 and 4 au, the planetesimals that were originally at a greater distance come closer to Mars and Earth and the originally colder planetesimals experience greater evaporation (blue vs. purple curves in Fig. \ref{fig:aini}).

The planetesimals between 2:1 resonance and the Jovian planet ($0.6 a_{\rm p}-1 a_{\rm p}$) are displaced from their initial locations, whereas those ($\ga 100$ km) existing sufficiently within secular and 2:1 resonances rarely migrate.
In this case, the eccentricity is maintained around the forced eccentricity provided by the Jovian planet.
The timescale at which the 100-km planetesimal at approximately 3 au migrates by 1 au is in the order of $\sim 10^8$ yr.
In Figure \ref{fig:aini}, the planetesimals originating at 3 au (yellow) and 2.3 au (red) are calculated for $10^8$ yr, and the planetesimals originating at 3.5 au (purple) and 4.1 au (blue) are calculated for $3\times 10^7$ yr.
Our results indicate that the planetesimals starting from 2--3 au do not migrate in a disk depletion timescale.
Planetesimals beyond 4.8 au are easily ejected outside the Jovian orbit. 
However, if the planetesimals experience a strong gas drag, they often return to the inner orbit because of the drag force and evolve similar to the planetesimals initiating from 4 au.

Unlike the inner planetesimals, the planetesimals beyond the orbit of the Jovian planet are not excited because of the following two reasons.
(1) The slow migration due to weak gas drag tends to keep the planetesimals at their initial locations. 
(2) The gas drag causes the planetesimals to migrate inward but the planet scatters them outward.
In this situation, the planetesimals are easily trapped in the mean-motion resonances, where they are highly excited; however, the excited planetesimals are easily scattered {\it outward} by the Jovian planet. 
The scattered planetesimals may return because of the gas drag but they are easily scattered again.
Thus, the resonance crossing cannot enhance the excitation as in the case of the inner region.
As the scattered planetesimals maintain a pericenter near 5.2 au, their maximum relative velocity is less than $\la 5 {\rm \ kms^{-1}}$.

\subsection{Dependence on the gas density}
\label{subsec:diskrho}

The gas density of the disk affects the magnitude of the drag force and location of secular resonance.
Strength of the specific gas drag force is proportional to the gas density $\rho$ and inversely proportional to the planetesimal size $r_{\rm a}$. 
In this sense, the effect of changing the gas density is comparable to changing the planetesimal size and the strength of the gas drag.
Lower gas density causes weaker gas drag. 
However, as the density decreases, the location of secular resonance shifts to the inner location with higher disk density to correspond to the precession rate of the pericenter of the Jovian planet. 
Although the maximum eccentricity that the planetesimal reaches is slightly higher when the gas density is low, they almost nullify each other's effects (Fig. \ref{fig:rho100}).
Thus, the effect of the density difference is small as long as the planetesimals cross secular resonance.

The evolutions of ten 100-km planetesimals are shown in Figure \ref{fig:rho100}.
The gas density is proportional to the disk surface density $\Sigma_1$, as shown in Equation (\ref{eq:diskdensity}).
The gas surface densities from the left to the right panels are 10\%, 50\%, 100\%, and 150\% of the minimum-mass disk. 
The red broken lines show the locations of secular resonances at the corresponding disk densities. 
We show the results of four different initial locations of the planetesimals (solid blue circles). 
From top to bottom, the initial semi-major axes are $a_{\rm ini}=$4.1, 3.5, 3.0, and 2.3 au.
The planetesimals initiating beyond secular resonance (top left panels) can reach 5 ${\rm km s^{-1}}$, independent of the disk density in the considered range, whereas those beginning the migration from well inside secular resonance (bottom right panels) barely obtain the high relative velocities.
From the simulations of $a_{\rm ini}$=3.5 and 3.0 au, we can deduce that the passage of 2:1 resonance (3.3 au) is helpful but not essential to the formation of highly excited planetesimals. 
Although the planetesimals ($a_{\rm ini}<$3.5 au) pass secular resonance when within 3:1 resonance ($\Sigma_1=0.1\Sigma_{\rm MMSN}$ case) and the planetesimals that cross the red broken line (secular resonance) can be excited, the resulting excitations are not high.

The green circles show the maximum eccentricities of the 300-km planetesimals.
Ten simulations were performed for each case.
As the gas drag in this case is weaker, the maximum eccentricities tend to be slightly higher than that of the 100-km planetesimals.

\subsection{Dependence on Jovian parameters}
\label{subsec:jupiter}

Marzari \& Weidenschilling (2002) showed that the excitation of eccentricity based on 3:2 resonance crossing does not require an elliptical Jovian orbit, whereas 2:1 resonance trapping requires a small percentage of Jovian eccentricity.
Moreover, the noncircular orbit of a Jovian planet ($e_{\rm p} \ga 0.02$) enhances the orbit crossing rate of the planet. 

\begin{figure}\epsscale{1.0}
\plotone{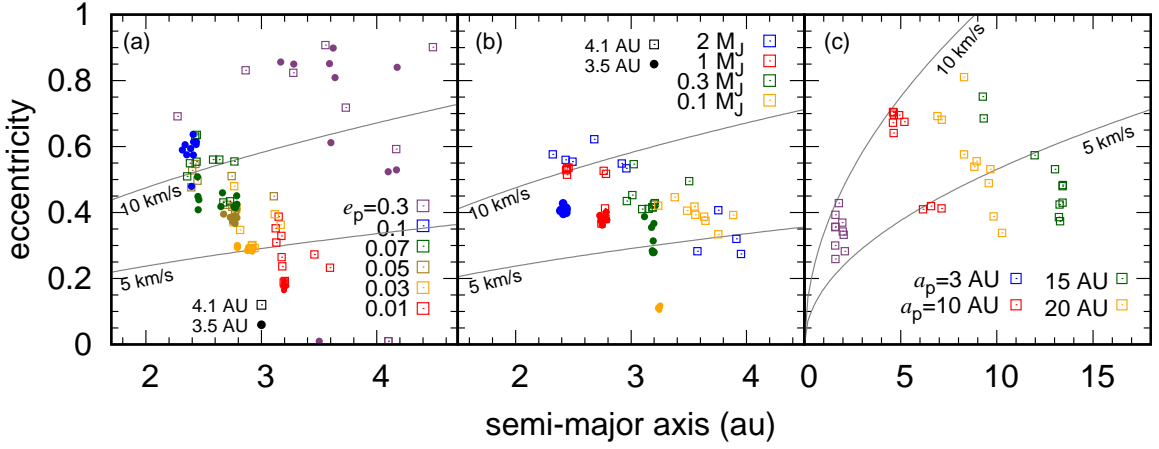}
\caption{ 
Maximum eccentricities of planetesimals with different Jovian planets.
Ten 100-km planetesimals were simulated for each parameter of the Jovian planet.
(a) Eccentricities of the planets are $e_{\rm p}=0.3$ (purple), $e_{\rm p}=0.1$ (blue), $e_{\rm p}=0.07$ (green), $e_{\rm p}=0.05$ (khaki), $e_{\rm p}=0.03$ (yellow), and $e_{\rm p}=0.01$ (red). The initial locations of the planetesimals are 4.1 au (squares) and 3.5 au (solid circles). 
(b) The masses of the planets are $2 M_{\rm J}$ (blue squares), $1 M_{\rm J}$ (red squares), $0.3 M_{\rm J}$ (green squares), and $0.1 M_{\rm J}$ (yellow squares). Initial locations of the planetesimals are 4.1 au (squares) and 3.5 au (solid circles). 
(c) The planet locations are 3 au (purple squares), 10 au (red squares), 15 au (yellow squares), and 20 au (green squares).
The initial semi-major axis of the planetesimals is $\sim 0.8 a_{\rm p}$.
(A colored version of this figure is available in the online journal.)
\label{fig:juppara}}
\end{figure}

As we newly consider secular resonance, which directly reflects the Jovian parameters, it would be interesting to study the dependence of the Jovian parameters on the orbits of the planetesimals.
Figure \ref{fig:juppara} shows the results of simulations in which different eccentricities, masses, and semi-major axes of planets were examined. 
Only one parameter was changed at a time in these simulations. 
For instance, when we changed the eccentricity, the mass and semi-major axis were kept at their current Jovian values.

The power of secular resonance is proportional to the eccentricity of the Jovian planet $e_{\rm p}$.
As the eccentricity of the Jovian planet increases, the maximum eccentricity that the planetesimal obtains increases (Fig. \ref{fig:juppara} $a$).
In this figure, the open squares and solid circles represent the maximum eccentricities of the planetesimals that were initially at 4.1 and 3.5 au, respectively. 
As shown in Figures \ref{fig:aini} and \ref{fig:rho100}, the maximum eccentricity of the inner planetesimal is slightly smaller, as the number of resonances it has passed through is smaller. 
However, the eccentricities at both planetesimal locations show a tendency to increase with the Jovian eccentricity.

The evolution for $e_{\rm p}=0.01$ is similar to the case without disk gravity.
For $e_{\rm p}=0.01$, approximately half the planetesimals at $a=4.1$ au can reach $e=0.3$ but the other half at  $a=4.1$ au and planetesimals at $a=3.5$ au cannot.
The Jovian planet in the gas disk would retain a small eccentricity; however, $e_{\rm p}\ga 0.03$ and  $e_{\rm p}\ga 0.05$ would be required to cause silicate crystallization and chondrule formation, respectively.
The maximum eccentricity for $e_{\rm p}>0.05$ is achieved at 3:1 resonance.
Therefore, although the eccentricity is enhanced using the Jovian eccentricity, the semi-major axis at the maximum eccentricity is always around $a \sim 2.5$ au (0.5$a_{\rm}$).
When the Jovian eccentricity exceeds 0.3, as is often observed in exoplanets, the scattering by the Jovian planet becomes strong.  This leads to maximum eccentricity of over 0.6.

Figure \ref{fig:juppara}$b$ shows the results with different planetary masses ($M_{\rm p}$).
As the planet mass decreases, the maximum eccentricity of the planetesimal with $M_{\rm p} < 1 M_{\rm J}$ decreases.
As the planetesimals cannot be excited inside 3:1 resonance, the peak eccentricity is maintained at approximately 0.5 even if the planet is more massive than the Jovian mass.
The location of secular resonance depends on the mass of the Jovian planet.
Secular resonance occurs at a more distant location when the planetary mass is small because the gravity of the planet needs to counterbalance the gravity of the disk.
When the planetary mass is 0.1 $M_{\rm J}$, 0.3 $M_{\rm J}$, 1 $M_{\rm J}$, and, 2 $M_{\rm J}$, secular resonance occurs at 4.2, 3.8, 3.2, and 2.9 au, respectively. 
When $M_{\rm p} = 0.1 M_{\rm J}$, the excitation is limited, as adequate number of planetesimals do not pass through secular resonance.
In addition, the planetesimals starting from 3.5 au cannot cross secular resonance with $M_{\rm p} < 0.5 M_{\rm J}$ and the eccentricity is not enhanced (the circular symbols at around 3.2 au). 
The depletion of planetesimals beyond 2:1 resonance becomes severe when the Jovian planet growth exceeds by more than 1/3 of the current Jovian mass. 
The planetesimals beyond 2:1 resonance start their migration at a relatively early stage of the Jovian growth, whereas the planetesimals within 2:1 resonance do not migrate until the mass of the Jovian planet increases sufficiently.
When $M_{\rm p} = 2 M_{\rm J}$, the strong scatterings remove approximately half the planetesimals before they reach secular resonance.
We tested a few cases by considering another planet (Saturn) in addition to Jupiter; the results scarcely changed as long as the location of secular resonance did not move considerably, as can be easily expected from the above-mentioned results. 

Figure \ref{fig:juppara}$c$ shows four simulations corresponding to different planet locations.
The purple, red, yellow, and green squares show the cases in which the semi-major axes of the Jovian planet are $a_{\rm p} =$3, 10, 15, and 20 au, respectively.
In each case, the maximum eccentricity is recorded near 3:1 resonance ($\sim 0.5 a_{\rm p}$).
This is because the evolution can be scaled by using Hill's radius apart from the effect of the disk, even if we change the planetary location.
We have shown in Figure \ref{fig:rho100} that the evolution of planetesimals is not sensitive to the disk gravity.
The gas drag is another affecting factor.
The maximum eccentricity tends to be reduced in the inner region, where the gas drag is effective ($\la$ 3 au).
However, the eccentricity can reach 0.3--0.4 even in the case of 3 au.
It suggests that the effect of the gas drag cannot be exactly scaled by Hill's radius; however, the damping is moderated because the mutual distance between the mean-motion resonances becomes shorter and the planetesimals enter the next resonance before the damping.
Note that even if the planetesimals have similar eccentricities, the duration for which the planetesimals retain high eccentricity is shorter and the relative velocity $v_{\rm rel}\sim ev_{\rm kep}$ is much faster in the inner region.

\vspace{10mm}

\section{IMPLICATION TO CHONDRULE FORMATION AND PLANETESIMAL EVAPORATION}
\label{sec:discussions}

\begin{figure}\epsscale{0.7}
\plotone{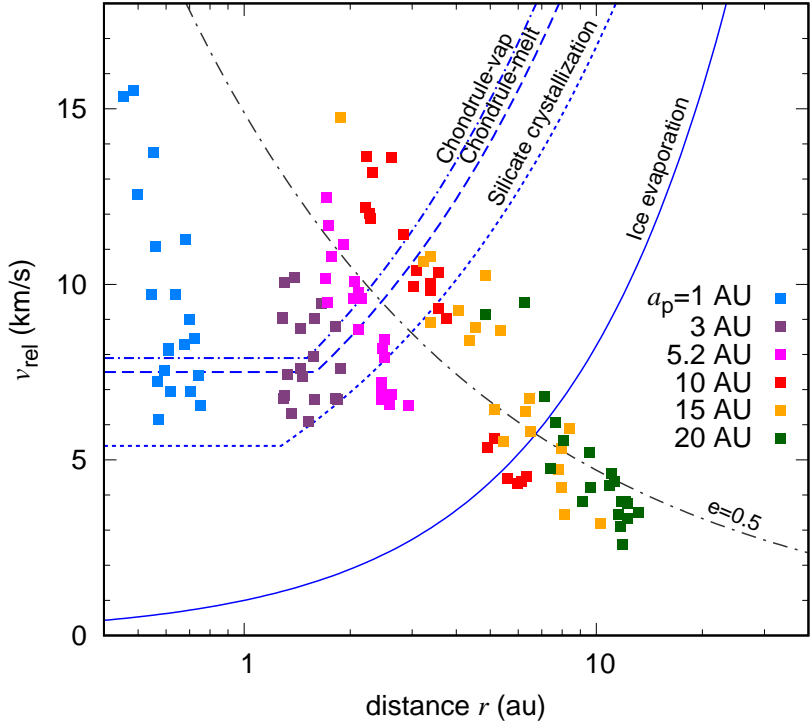}
\caption{
The maximum velocity of the planetesimals relative to the disk versus the location at which the velocity is recorded inside the planetary semi-major axis.
Six different semi-major axes of the Jovian planet were examined.
Twenty 100-km planetesimals were simulated for each planetary location.
The initial locations of the planetesimals were set to approximately 80\% of the location of the planet.
Gray dashed-dotted line shows the maximum relative velocity of the planetesimals with $e=0.5$.
The required shock speeds for chondrule melting (1900 K), chondrule vaporization (2100 K), and crystallization of amorphous silicate (1000 K) are shown by blue broken line, blue dashed-dotted line, and blue dotted line, respectively. 
The required velocity for the evaporation of icy planetesimal with Stanton number 0.05 is shown by a solid line. 
(A colored version of this figure is available in the online journal.)
\label{fig:condition}}
\end{figure}

The highly excited planetesimals can be a source of bow shock, which not only heats up the surrounding dust particles (e.g., Hood 1998) but also causes the evaporation of icy planetesimals (Tanaka et al. 2013).
Figure \ref{fig:condition} shows the maximum relative velocity that we obtain from various initial locations of the planet ($a_{\rm p}$) together with the shock velocity required to cause evaporation and other thermal effects. 
In this plot, simulations were performed for the density of the minimum-mass disk, 100-km sized planetesimals, and the Jovian planet mass.
When the Jovian planet is located beyond 5 au, the typical maximum velocity is approximately $e V_{\rm kep, r}= 0.5 V_{\rm kep, r} $ (the line marked as $e=0.5$).
If the planetesimals are excited enough, the top velocity occurs at $r \sim 0.5 a_{\rm p}$ (3:1 resonance), i.e., the typical relative velocity is $v_{\rm rel} \sim 9 (a_{\rm p}/{\rm 5 \ au})^{-1/2} {\rm kms^{-1}}$ ($a_{\rm}>5$ au).
In the orbit of the peak velocity, the relative velocity varies between $\sim 1/2 ev_{\rm kep}$ and $\sim ev_{\rm kep}$ (see \S \ref{subsec:velocity}) during one Kepler orbit and the planetesimal travels from $1/4 a_{\rm p}$ to $3/4 a_{\rm p}$.

Jovian resonances are proposed as the main mechanism that excites the planetesimals to cause the shock waves (Weidenschilling, Marzari, \& Hood 1998; Marzari \& Weidenschilling 2002; Nagasawa et al. 2014).
Experimental studies and meteorite analysis constrain the chondrule formation models to short heating time ($\sim$ minutes), peak temperature of $\sim $ 2000 K, short cooling time ($\sim$ hours), and high oxygen fugacity (e.g., Jones et al. 2000 and references therein), and the shock heating satisfies most of these constraints (e.g., Hood 1998; Iida et al. 2001; Desch \& Connolly 2002; Ciesla, Hood, \& Weidenschilling 2004; Boley, Morris, \& Desch 2013).
We can derive the required velocity ($v_{\rm melt}$) for chondrule melting (1900 K) by using equation (32) of Iida et al. (2001).
The melting of the chondrule precursors occurs in the region over the broken lines in Figure \ref{fig:condition}. 
{ Similarly, a blue dashed-dotted line shows the condition for initiation of chondrule vaporization (2100 K), as obtained from equation (35) of Iida et al. (2001). 
Note that the formation of chondrules requires the maximum speed of the planetesimal to be higher than that required for the melting condition, but it may not necessarily lie between the lines of melting and vaporization conditions. 
Even if it is above the vaporization line, the planetesimal can contribute toward the formation of chondrules because the moderate bow-shoch-heating condition will be realized at places distant from the center of the planetesimal (Hood 1998; Ciesla et al. 2004).
In addition, the figure shows the maximum velocities of the planetesimals, i.e., planetesimals have smaller relative velocity in the wide region during the evolution (see Fig. \ref{fig:vrelecc}). 
Thus, the planetesimal, which evaporates the chondrule precursors in the narrow region near the peak speed, can contribute toward the melting of chondrule precursors in the wide regions.
The chondrules shrink in radii in the region above the line of vaporization, but it does not imply total evaporation (Miura, Nakamoto, \& Susa 2002).}
More than half of the planetesimals can contribute toward chondrule formation when the Jovian planet is formed within $\sim$ 5 au. 
Chondrule formation is also possible for $a_{\rm p}=10$ au, but in this case, the mass of the current Jupiter is required.
As the relative velocity is maintained at $v_{\rm rel} \la 6$ ${\rm kms^{-1}}$, the chondrule formation by the 10-km planetesimal bow shock alone is difficult. 
In larger planetesimals, the melting of chondrule precursors would be possible with $M_{\rm p} \ga 0.3 \ M_{\rm J}$ as long as we consider orbital parameters similar to Jupiter.
The current Jovian eccentricity, $e_{\rm p}\sim 0.05$, can satisfactorily account for chondrule formation by using our mechanism. 
Eccentricity smaller than 0.03 (see Fig. \ref{fig:juppara}$a$) would not be sufficient to form chondrules in the asteroid regions.

The solid line in Figure \ref{fig:condition} shows the velocity that results in the evaporation of icy planetesimals obtained using equation (20) by Tanaka et al. (2013), in which the Stanton number expressing the efficiency of heat conduction is taken as 0.05.
The evaporation line and $e=0.5$ line cross at approximately $r \sim 7$ au, i.e., $a_{\rm p}\sim 14$ au.
We find that the evaporations of icy planetesimals occur when a Jovian planet is formed within $\sim$15 au and grows larger than $M_{\rm p} \ga 0.1 M_{\rm J}$ (see Fig. \ref{fig:juppara}$b$). 
The figures show that the icy planetesimals evaporate efficiently even beyond the snow line.
{ The weak shock waves cause evaporation of icy planetesimals, although chondrule formation does not occur. 
The evaporation of planetesimals continues during the high eccentricities of the planetesimals with a timescale of approximately $10^5-10^6$ yr. 
Consequently, icy planetesimals change to rocky ones with less abundance of water.
High speed planetesimals also cause sublimation nearby free-floating ice. 
Although the amount of released water depends on the number density of planetesimals and remaining icy materials in the protoplanetary disk, these processes indicate that a huge amount of water is released to the protoplanetary disk.  
The evaporation of icy planetesimals may also change the gas components in the protoplanetary disk because of chemical reactions with other evaporated volatile molecules except water, as shown later.  
On the other hand, the strong shock waves satisfying the condition for chondrule formation cause both the processes of chondrule formation and icy planetesimal evaporation because the evaporation condition is fully satisfied, as shown in Figure \ref{fig:condition}.  
Accordingly, the water vapor increases in the post-shock region around the planetesimal, where the chondrule precursors melt. 
This effect can cause high oxygen fugacity, which is consistent with estimated chondrule formation condition.}

The planetesimal bow shock heats the surrounding gas at $\sim$1000 s (Iida et al. 2001). In that timescale, crystallization occurs at $\sim$1000 K (Tanaka, Yamamoto, \& Kimura 2010).
By using the formula (32) proposed by Iida et al. (2001), we can obtain the condition for which the shock heating causes crystallization of amorphous silicate dust.
The condition is represented by a dotted line in Figure \ref{fig:condition}.
As shown, the heating due to shock is enough to crystallize the amorphous silicate, provided the Jovian planet is formed within $\sim$10 au.

Secular resonance is important to excite the planetesimals enough to cause crystallization and chondrule formation.
Without secular resonance, the maximum eccentricity is typically $\la$0.4 near 2:1 resonance, i.e., more massive or more eccentric Jovian planet is required to cause crystallization and chondrule formation. 
Although the location of secular resonance and magnitude of gas drag are dependent on the gas disk, they can be evaluated analytically. 
We can estimate the maximum velocity even in other disk systems based on the analogy of this work.
The dynamical effect on a planetesimal is less dependent on the density of the gas disk $\rho$ except for the migration timescale; however, the temperature of the chondrule precursor heated owing to the bow shock increases with $\rho$ (e.g., Hood 1998; Iida et al. 2001; Desch \& Connolly 2002). 
Thus, the melting of the chondrule precursor for a disk with low gas density requires high relative velocity; nevertheless, the maximum relative velocity barely changes as long as we consider the current eccentricity of Jupiter.

\begin{figure}\epsscale{0.7}
\plotone{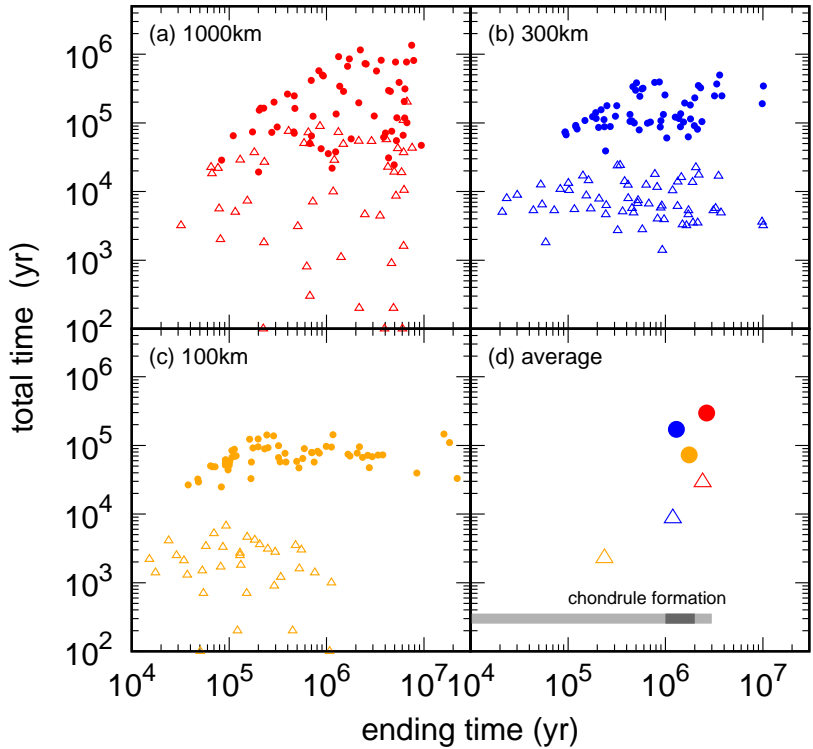}
\caption{ 
Total duration and ending time of the period wherein the planetesimals have evaporation-speed (circles) and melting-speed of chondrules (triangles). 
Panels ($a$)-($c$) show the data for different planetesimal sizes (100--1000 km).
Planetesimal sizes are shown in each panel.
Panel ($d$) shows their average values.
Planetesimal radii are 100 km (yellow), 300 km (blue), and 1000 km (red).
Initial locations of the planetesimals are $a=$4.1 au, the disk surface densities are that of the minimum-mass disk, and current Jovian parameters are assumed.
(A colored version of this figure is available in the online journal.)
\label{fig:time}}
\end{figure}

Circles in Figure \ref{fig:time} show the total estimated time (cumulative time) for which each planetesimal has sufficiently high speed for evaporation and the end time of the duration when the planetesimal recorded that velocity.
The time was measured for sixty planetesimals for each case of 1000 km (panel $a$), 300 km (panel $b$), and 100 km (panel $c$) sizes.
We used the Stanton number to express the efficiency of heat conduction of $\alpha=0.05$ (Tanaka et al. 2013) to estimate the critical velocity for evaporation at that distance. 
Note that the total time calculated here is an approximate value, as the velocity is checked every 100 yr to add the time by which the relative velocity exceeds the critical velocity. 
The circular symbols in panel $d$ show the average values of the planetesimals, which exceed the evaporation speed. 
The planetesimals evaporate at $\sim 10^5$ yr $-10^6$ yr, and this evaporation event continues for several Myr.

The 1000-km planetesimals might be considered as less efficient than the 300-km planetesimals because more than half of the planetesimals were ejected out of the Jovian orbit.
At the end of our simulation, 6 out of 60 planetesimals (300 km) and none out of 60 planetesimals (100 km) were ejected outside. 
However, the 1000-km planetesimals remained for a longer time at 2:1 resonance with $e \ga 0.2$.
This implies that, as shown in Sec. \ref {subsec:velocity}, they cross the asteroid region with high speed for a long time, which is generally sufficient for the longer evaporation time to be recorded.

When the icy planetesimal in the asteroid region of the minimum-mass gas disk has an eccentricity of 0.1--0.2, its size is reduced to half in $10^7$ yr because of bow shock (Tanaka et al. 2013).
Our results show that the timescales in which the 100- and 1000-km planetesimals have an evaporation speed of $\sim$ 3 ${\rm kms^{-1}}$ are $\sim 10^5$ yr and a small multiple of $10^5$ yr, respectively.
This time is less than $10^4$ yr for 10 km-planetesimals.
As pointed out by Marzari \& Weidenschilling (2002), the time for which an individual planetesimal experiences shock heating is not long.
In addition, an individual icy planetesimal would not completely dry up.
However, overall, the large velocity relative to the gas disk causes a non-negligible amount of evaporated materials even beyond the snowline.

In our solar system with the current location of Jupiter, icy planetesimals beyond the main belt of asteroids migrate to the terrestrial region. 
If they retain water, Earth would gain a considerably large amount of water.
The evaporation of planetesimals is important to reduce the water fraction because the planetesimals, which reach the Earth's orbit, experience very high velocity.
The estimation of water delivery and distribution of materials would be studied in the future.

When we assume the minimum-mass disk, the total mass of the planetesimals between 3 and 4 au is $2 \times 10^{28}$ g. 
If all planetesimals have 100-km sizes, their total number would be $\sim 5 \times 10^{6}$.
In our simulations, the planetesimals retain high speed for $10^5$ yr in the evaporation time of $10^6$ yr, i.e., 10\% of the planetesimals are highly excited.
This implies that approximately 500,000 planetesimals at the maximum are excited simultaneously. 
Of course, there would be a variation in size and the majority of the mass would be of smaller-sized planetesimals; however, a non-negligible number of planetesimals causes shocks between $\sim 1/4 a_{\rm p}$ to $\sim 3/4 a_{\rm p}$.

The evaporation of icy planetesimals ejects molecules of ice in a protoplanetary disk.
The chemical evolutions in stationary disks, whose temperatures do not change considerably, have been well studied (e.g., Walsh et al. 2014; Furuya \& Aikawa 2014). 
The chemical reactions in the gas-phase induce the evaporation of icy planetesimals owing to the enhancement of the local temperature, and daughter molecules are formed. 
These parent molecules evaporate from ice and their daughter molecules are observed near comets (e.g., Mumma \& Charnley 2011). 
Some of them, such as ${\rm H_2S}$ and SO, are known as shock tracers and are not observed in the stationary T Taurus disks. 
Once icy planetesimals evaporate, they would become observable.
Accordingly, in extrasolar systems, the depletion of the planetesimals between a Jovian planet and its resonance through evaporation can be observed by ALMA if the evaluation is conducted beyond 5--10 au and continues for a disk's lifetime. 
Fractional abundance of ${\rm H_2S}$ and SO relative to hydrogen nuclei greater than $\sim 10^{-10}$ is required for detection within reasonable observation time using the ALMA.
In a simple estimation, fractional abundance will be achieved when approximately 0.01\%--0.1\% of the planetesimals at the location evaporate, i.e., 1000 planetesimals evaporate every $10^4 -10^5$ yr (Nomura et al. 2018, in prep.). 

It is preferable that the line emitting region is large (beyond $\sim$ 5 au) in order to detect the line emission by the ALMA observations, whereas it becomes difficult for icy planetesimals to evaporate in the outer region, as is shown in Figure \ref{fig:condition}.
In our simulations, when a Jovian-sized planet was formed at 15 au, approximately 3/4 of the planetesimals reached the evaporation speed. 
Individual planetesimals evaporate over $\sim 10^5$ yr, and the evaporation typically continues for $\sim 10^7$ yr. 
Therefore, roughly 1\% of the planetesimals initially between 9 and 15 au evaporate simultaneously at approximately 4--10 au.
We suppose that the observation of planetesimal evaporation would be possible, especially in the case of a massive disk holding many planetesimals.

Note that the location of the evaporation is not near the planet.
In the case of minimum-mass disks, secular resonance appears between 2:1 and 3:1 resonances, and the maximum eccentricity of the planetesimals is achieved near 3:1 resonance (Fig. \ref{fig:reso}). 
The location of secular resonance depends on the disk density and gravity of other planets.
Therefore, the high-speed region of the planetesimals can be far from the Jovian planet.
The location of evaporation of materials could provide us with information on the gas density and planetary mass. 
As the high-speed collisions would cause formation of dust materials, the dust ring covering the resonant region might also be observable.

Isotopic analyses show that most chondrules are formed during 1--3 Myr after the formation of Ca-Al-rich inclusions (CAIs) (e.g., Kita \& Ushikubo 2012).
The triangles in Figure \ref{fig:time}$d$ show the average time during which the planetesimals have enough speed to cause chondrule melting.
As the required velocity is a more difficult condition than evaporation, the total time tends to be shorter.
For example, in the case of 300-km planetesimals, 59 of 60 planetesimals experience $v_{\rm rel} \ga v_{\rm melt}$, and shock generating events typically continue until $\sim 10^6$ yr after the Jovian formation. 
The planetesimals with radii larger than 300 km have sufficient high-speed periods to explain the duration of chondrule formation. 
Although the standard deviation is almost of the same order as the average value, a rough trend is observed, in which the time increases with the planetesimal size. 
The shock generating duration for 1000-km planetesimals is slightly longer than that for 300-km planetesimals.
The 100-km planetesimals are less efficient, as the individual excitation time is approximately ten times shorter; however, they contribute toward chondrule formation in the early stages.
We also expect that the total number of planetesimals in a size range increases as the size decreases.
Note that the duration we have shown in this study is the time after the formation of Jupiter and not the time after the formation of CAIs.
If Jupiter is formed approximately 1 Myr after the CAI formation, the required duration of planetesimal shock should be 0 yr to 2 Myr in our figure. 
In addition, note that we neglected the disk depletion in our simulations. 
When the disk is depleted, the shock is no longer generated.
Our results fit the timing suggested by the isotopic data if Jupiter was formed around 1 Myr after the CAI formation and the disk was dissipated approximately 2 Myr after Jupiter formation.
We can conclude that icy planetesimals ($\ga $100 km) experiencing secular resonance beyond 3:1 resonance can evaporate and become shock sources of chondrule formation in the disks similar to the minimum-mass disk.

In this study, we neglected the effect of precession of the gas disk due to the Jovian planet.
If we consider that the gas disk is also precessed because of the Jovian planet, the relative velocity is not high at the location of secular resonance. 
However, the velocity that is important to us is not at the location of secular resonance but at the location of 3:1 resonance.
The excitation of the eccentricity at the location of secular resonance is important because the planetesimals are pushed toward 3:1 resonance.

In our simulations, the evolution of planetesimals was determined based on the balance of the damping due to the gas drag and excitation due to a Jovian planet.
The balance of the two effects could be changed by other effects.
For example, dynamical friction in a system with protoplanets excites the planetesimals. In addition, perturbation from the protoplanet drops the planetesimals from the resonance (Hood \& Weidenschilling 2012).
In the simulations by Hood \& Weidenshilling (2012), in which collisions and close encounters with other planetesimals were considered, only bodies as large as half the Moon or larger could possibly attain eccentricities of $\ga$ 0.4.
We did not consider size change due to the evaporation and the effects of the collisions.
These effects should be considered in the next step.
The generation of shock is continued during the lifetime of the disk.
Nevertheless, we neglected the effect of the depletion of the disk.
During the depletion of the disk, secular resonance migrates and sweeps the asteroid region.
Although the sweeping of secular resonance would not be important in the later stage of gas depletion, the effect of the sweeping should be examined, as the disk gas density of approximately more than 1/3 of minimum-mass disk would be required to generate shock (Tanaka et al. 2013).

\section{CONCLUSIONS}
\label{sec:conclusions}

We studied the evolution of planetesimals embedded in the gas disk after a Jovian planet is formed.
The planetesimals migrate inward crossing planetary resonances because of gas drag.
The eccentricity of 100-km planetesimals is enhanced typically up to $e\sim 0.4-0.6$. 
The excited planetesimals are initially located between the planet and 2:1 resonance, which exists at approximately $\sim 60$\% of the planetary semi-major axis (0.6$a_{\rm p}$). 
We found that the peak eccentricity is normally achieved at the location of 3:1 resonance, which exists at approximately $\sim 0.5 a_{\rm p}$.
The relative velocity with respect to the gas was reached at $\sim$10 ${\rm kms^{-1}}$, when the planet had the current orbital parameters of Jupiter.  

We performed simulations for various parameters of the planetesimals, the Jovian planet, and the gas disk. 
The results from these simulations are summarized as follows:

1) {\it Planetesimal size}: 
The larger planetesimals (with 300-1000-km radii) are more excited, as the gas drag becomes less effective (Fig.\ref{fig:size}). The larger the planetesimals, the longer the evolution time of the planetesimals. The higher eccentricity causes longer migration distances until the circularization.

2) {\it Planetesimal location}: 
The planetesimals between 2:1 resonance and the planet are the most excited (Fig. \ref{fig:aini}). The planetesimals with semi-major axes smaller than that of secular resonance are scarcely excited (Figs. \ref{fig:aini} and \ref{fig:rho100}).

3) {\it Eccentricity of the Jovian planet}: Non-zero eccentricity is required for excitation of the planetesimals. Although higher eccentricity is favorable for higher excitation (Fig. \ref{fig:juppara}$a$), when the eccentricity of the Jovian planet exceeds 0.3, the planetesimals are scattered rather than contributing to the source of the shock wave in the asteroid region.

4) {\it Mass of the Jovian planet}: A massive planet causes greater excitation of the planetesimals (Fig. \ref{fig:juppara}$b$). 
However, there is a limit to the excitation at approximately 3:1 resonance. 
In our solar system, the planetesimals beyond 4 au start to generate bow shocks when Jupiter grows larger than 0.1 $M_{\rm J}$. The planetesimals at approximately 3.5 au need to wait until Jupiter grows to its current mass.

5){\it Location of the Jovian planet}: The dependence on the magnitude of the excited eccentricity is weak (Fig. \ref{fig:juppara}$c$); however, the relative velocity decreases as the planetary semi-major axis increases because of the Kepler velocity (Fig. \ref{fig:condition}).

6){\it Disk density}: The location of secular resonance depends on the disk density. 
The density has only a small impact as long as the planetesimals cross secular resonance (Fig. \ref{fig:rho100}).

When Jovian planets grow in the gas disks, the planetesimals migrate causing bow shocks.
In our solar system, the process of planetesimal bow shock would be important to connect the formation of giant planets to the chronological and chemical data from meteorites. 
We also find that:

7) The excitation due to secular resonance of the Jovian planet is enough to cause melting of the chondrule precursors and crystallization of silicate for the wide range of Jovian locations less than 10 au (Fig. \ref{fig:condition}).
The duration where supersonic planetesimals ($\ga$ 100-km radii) exist continues over $\ga$1 Myr after the formation of the Jovian planet (Fig. \ref{fig:time}). 
It would be consistent with the timescale of chondrule formation suggested by the isotopic data (1--3 Myr).

8) The excitation causes the evaporation of icy planetesimals ($\ga$ 100-km radii)  owing to the bow shock heating for a wide range of parameters that we tested, even beyond the snow line (Fig. \ref{fig:condition}).
The conditions we found are not rare in the exosystems. 
Therefore, the evaporation beyond the snow line is an interesting target of observations because it gives a new clue to the formation of giant planets.

\acknowledgments 
We are thankful for a careful and helpful review by an anonymous referee.
This work was supported by JSPS KAKENHI(17K05642, 16H04041, 16H00927, 26287101).



\end{document}